\documentclass[aps,pre,superscriptaddress,noshowpacs]{revtex4}
\usepackage[intlimits]{amsmath}
\usepackage{amsfonts}
\usepackage{psfrag}
\usepackage{graphicx}
\usepackage{subfigure}
\usepackage[usenames]{color}
\usepackage{indentfirst}
\usepackage{ifpdf}

\ifpdf
\usepackage{epstopdf}
\usepackage[pdftex]{hyperref}
\else
\usepackage[hypertex,ps2pdf,colorlinks,urlcolor=blue,citecolor=blue]{hyperref}
\fi

\advance\voffset by -1.4cm
\advance\hoffset by -0.3cm
\newcommand{\bigx}[1]{\bBigg@{#1}}

\def\3pt#1#2#3{{\langle{#1}\vert{#2}\vert{#3}\rangle}}

\newcommand\doi[2]        {\href{http://dx.doi.org/#1}{#2}}

\newcommand{\beq}{\begin{equation}}
\newcommand{\eneq}{\end{equation}}
\newcommand{\bdis}{\begin{displaymath}}
\newcommand{\edis}{\end{displaymath}}

\begin{document}

\title{Modulational instabilities in lattices with power-law hoppings 
and interactions}

\author{Giacomo Gori}
\affiliation{ICTP, Abdus Salam International Centre for Theoretical 
Physics, Strada Costiera 11, 34151 Trieste, Italy}

\author{Tommaso Macr\`i}
\affiliation{Max Planck Institute for the Physics of Complex Systems, 01187 Dresden, Germany}

\author{Andrea Trombettoni}
\affiliation{CNR-IOM DEMOCRITOS Simulation Center and SISSA, 
Via Bonomea 265 I-34136 Trieste, Italy \& \\ 
INFN, Sezione di Trieste, I-34127 Trieste, Italy}

\begin{abstract}
We study the occurrence of modulational instabilities in lattices 
with non-local, power-law hoppings and interactions. 
Choosing as a case study the 
discrete nonlinear Schr\"odinger equation, we consider 
one-dimensional chains with power-law decaying interactions 
(with exponent $\alpha$) and hoppings (with exponent $\beta$): An extensive 
energy is obtained for $\alpha, \beta>1$. 
We show that the effect of power-law interactions 
is that of shifting the onset of the modulational instabilities 
region for $\alpha>1$. At a critical value of the interaction strength, 
the modulational stable region shrinks to zero.
Similar results are found for effectively short-range nonlocal
hoppings ($\beta > 2$): At variance, for longer-ranged hoppings ($1 <
\beta < 2$) there is no longer any modulational stability. The hopping
instability arises for $q = 0$ perturbations, thus the system is most
sensitive to the perturbations of the order of the system's size. We
also discuss the stability regions in the presence of the interplay
between competing interactions - (e.g., attractive local and repulsive
nonlocal interactions). We find that noncompeting nonlocal
interactions give rise to a modulational instability emerging for a
perturbing wave vector $q = \pi$ while competing nonlocal interactions
may induce a modulational instability for a perturbing wave vector $0
< q < \pi$. Since for $\alpha > 1$ and $\beta > 2$ the effects are
similar to the effect produced on the stability phase diagram by
finite range interactions and/or hoppings, we conclude that the
modulational instability is ``genuinely'' long-ranged for $1 < \beta <
2$ nonlocal hoppings.
\end{abstract}
\maketitle

\section{Introduction}
\label{introd}

The investigation of the effects of the interplay between discreteness and 
nonlinearity is a long-standing argument of research in the study of the 
dynamical properties of nonlinear lattice models 
\cite{flach98,braun98,hennig99,ablowitz04,campbell04,malomed06,flach08}. 
A typical feature exhibited by nonlinear classical Hamiltonian lattices
is the existence of discrete breathers, i.e. time-periodic and 
space-localized solutions of the equations of motion. The study of their 
dynamical stability, as well as 
their robustness in long transient processes and thermal
equilibrium, has been the subject of an intense experimental and 
theoretical work \cite{flach08}. 

The interplay between discreteness and nonlinearity is also
crucial for the occurrence of modulational
instabilities (MI), well known in the theory of nonlinear media 
\cite{flach98,braun98}. MI are dynamical instabilities characterized by
an exponential growth of arbitrarily small fluctuations resulting from 
the combined effect of dispersion
and nonlinearity. The occurrence of modulational instabilities has been 
studied in a number of physical systems, ranging from fluid dynamics 
\cite{benjamin67} to nonlinear optics \cite{agrawal07}. The role and the 
consequences of the MI in the dynamics of discrete systems 
have been extensively studied: The MI was discussed in the context
of the discrete nonlinear Schr\"odinger equation (DNLSE) \cite{kivshar92}, 
which is a paradigmatic lattice model \cite{kevrekidis01} used 
to study nonlinear discrete dynamics \cite{hennig99,ablowitz04}. The DNLSE 
is commonly used to describe the effective dynamics in different physical 
systems of interest, including the dynamics of ultracold atoms in optical 
lattices \cite{trombettoni01} and optical waveguide arrays 
\cite{eisenberg98}. For ultracold bosons in optical lattices the onset 
of MI was analytically predicted \cite{smerzi02} and experimentally observed 
\cite{cataliotti03}, and in nonlinear waveguide arrays 
the experimental observation of the MI was also reported \cite{meier04}.

In this paper we study the occurrence of MI in the 
DNLSE in the presence of non-local long-range hoppings and interactions. 
A motivation for such a study comes from experiments 
with ultracold dipolar bosonic gases \cite{lahaye09,trefzger11} 
which have been Bose condensed recently by several groups \cite{griesmaier05,bismut10,lu11}, 
from the attainment of quantum degeneration for ensembles of polar molecules 
\cite{ospelkaus10}, and from the recent experimental investigations 
of strongly interacting Rydberg gases \cite{heidemann08, schauss12, saffman10}. 
Since the interaction potential 
in (di)polar gases decay as a power law $1/r^3$ 
(for Rydberg gases interacting through van der Waals interactions as $1/r^6$), 
recent experiments with dipolar gases in optical lattices 
\cite{muller11,billy12,depaz12}, 
as well as the realization of long-lived dipolar molecules 
in a three-dimensional periodic potential 
\cite{chotia11} and in perspective the dynamics of Rydberg atoms 
in optical lattices
\cite{henkel10,viteau12}, open the possibility to study DNLSE with non-local 
interactions.

Our other motivation is related to the wide interest in systems with 
long-range interactions \cite{leshouches10}. 
In these systems the range of interaction 
of the constitutive units is not bounded. A typical form of interactions,
relevant for a number of systems ranging from gravitational ones to
dipolar magnets and gases, is provided by the power-law decay $1/r^\gamma$
(e.g. for gravitational systems $\gamma=1$) where $r$ is the distance 
among the constituents. For statistical mechanics models, like the Ising or 
more generally the $O(n)$ models, the possibility to have power-law 
couplings makes possible the appearance of a rich phase diagram 
\cite{fisher72,sak73,luijten02}.

A first criterion to determine the long-rangedness of a system with power-law
decaying interaction is the comparison with the dimension $d$ of the space; 
as $\gamma$ is smaller than or equal to $d$, 
if the system is homogeneous and the interaction favours
homogeneity we obtain a diverging energy
density, thus to obtain a well
defined thermodynamic limit (if relevant) a rescaling
of the energy is in order (the so called Kac rescaling) 
\cite{ruffo09}. In the following we will refer to this region as 
the non-extensive long-range region.
If $\gamma$ is larger than $d$ the energy of the system is 
extensive and it is normal to individuate a value of $\gamma$, 
which we denote by $\gamma^\ast$, such that for $\gamma>\gamma^\ast$ 
the system behaves as a short-range system. Since it is $\gamma^\ast>d$, 
there is a region of values of $\gamma$, given by 
$d<\gamma<\gamma^\ast$, in which the behavior of the system 
significantly differs from the properties of the same system with short-range
interactions, although the energy is extensive. Such a region is the extensive 
long-range region, also referred to as the 
weak-long-range region \cite{ruffo09}, 
where ``weak'' refers to the extensivity of the energy. The actual value of 
$\gamma^\ast$ depends on the specific model and the dimension: For $O(n)$ 
models in $d=1$ it is $\gamma^\ast=2$ \cite{kosterlitz76, spohn99}. 

Both the thermodynamics and the dynamics of long-range interacting systems 
are extremely interesting \cite{leshouches10,ruffo09}. In particular, 
in the long-range region the dynamical evolution evidences that the system may 
stay in a quasi-stationary metastable state 
(different from the thermal equilibrium one) 
for a time exponentially growing with the size of the system. Such a metastable 
state is reached after a short-time dynamics, 
referred to as violent relaxation \cite{ruffo09}. 

While the main bodies of the studies on the dynamics of nonlinear lattices 
have dealt with short-ranged systems, the extensions of these results to 
long-ranged systems appeared in the literature 
addressing the properties of discrete
systems with different kinds of non-local dispersion or non-local 
nonlinear interaction we mentioned
\cite{gaididei97,flach98_2,christiansen98,mingaleev00,fratalocchi05}
and focused on the existence and stability of localized excitations 
(for a recent review on nonlinear waves in lattices  
see \cite{kevrekidis11}).

The purpose of the present paper is to study how modulational instabilities 
emerge in nonlinear lattices with non-local interactions and hoppings, 
aiming both at unveiling if (and in which conditions) short time  
dynamical instabilities occur in nonlinear lattices and at clarifying 
the nature of the emerging modulational instability.  We choose 
the DNLSE as a case study not only due to its paradigmatic usefulness, but 
also due to its relation with $XY$ [i.e., $O(2)$] models: 
When the fluctuation of the number of particles are frozen, the kinetic 
term in the DNLSE energy is basically the $XY$ model 
(see the discussion in Sec. 
\ref{the_model}). This is the reason why we choose to consider not only 
power-law interactions (as it is relevant for experiments with ultracold 
dipolar bosons in optical lattices), but also power-law hoppings [which 
corresponds to power-law couplings in $O(2)$ models]. Using the 
DNLSE we study the modulationally 
stable and unstable regions in the presence of power-law non-local interactions 
and hoppings, discussing also the interplay between local and 
non-local interactions, e.g. local attraction and non-local repulsion. 

The plan of the paper is as follows. In Sec. \ref{the_model}, 
we introduce the DNLSE with long-range hoppings and interactions
and discuss its relation with other statistical mechanics models. 
In Sec. \ref{MI_analysis} we derive the Bogoliubov spectrum of elementary excitations,
present the general framework for the determination of the stability regions 
in presence of power-law interactions and hoppings and specialize it
to the analysis to the short-range limit reminding 
the known results of the MI analysis \cite{kivshar92,smerzi02}.  
Our findings for power-law interactions are presented in 
Sec. \ref{MI_LR_interaction} where we also consider the case of 
attractive and competing local and non-local interactions.
Sec. \ref{MI_LR_hopping} deals with a system with non-local hopping
and local interaction which presents some peculiar features
with respect to the long-range interaction which are further
investigated in \ref{MI_interaction_hopping_comparison}.
The physical applications of our results, in particular to ultracold dipolar 
gases in optical lattices, are discussed in Sec. \ref{phys_app}, while our conclusions are in Sec. \ref{conclusions}.

\section{The DNLSE with long-range interactions and hoppings} 
\label{the_model}

The DNLSE with non-local interactions and hoppings 
reads
\begin{equation}
i \hbar \frac{\partial \psi_j}{\partial \tau} = - \sum_{m} 
t_{j,m} \psi_m + \sum_m V_{j,m} \mid \psi_m \mid ^2 \psi_j.
\label{DNLSE}
\end{equation}
In Eq. \eqref{DNLSE} $\tau$ is the time and 
the indices $j,m$ denote the sites of a lattice. For simplicity we assume 
that the lattice is one dimensional, but the subsequent 
analysis can be extended to higher dimensional lattices. 
The indices $j, m$ then 
assume the values $j, m=0,\ldots,L-1$ 
($L$ is the number of the sites, 
taken to be even). Periodic 
boundary conditions will be also assumed, so that the wavefunction satisfies
the condition $\psi_j=\psi_{j+L}$.
The Hamiltonian corresponding to Eq.~\eqref{DNLSE} reads
\begin{equation}
H_{\text{DNLSE}}=-\sum_{j,m} \psi_j^\ast t_{j,m} \psi_m +\frac{1}{2} \sum_{j,m} \mid \psi_j \mid^2 V_{j,m} \mid \psi_m \mid^2.
\label{def_H}
\end{equation}
We denote the diagonal interaction by
\begin{equation}
V_{j,j} = U,
\label{diag_U}
\end{equation}
and the next-neighbor interaction as
\begin{equation}
V_{j,j+1} = V.
\label{non_diag_V}
\end{equation}
The interaction coefficients $V_{j,m}$ in Eq.~\eqref{DNLSE} 
are assumed 
to be power-law decaying with exponent $\alpha$, i.e. 
$\sim 1/\mid m-j \mid^\alpha$. Since $V_{m,j}=V_{j,m}$, implementing 
the periodic boundary conditions amounts to require that 
$V_{j,m}=V_{\left( j+n \right) \mod L, 
\left( m+n \right) \mod L}$, where $j \le m$ and $n=1,\ldots,L-1$. 
We have therefore
\begin{equation}\label{VV}
V_{0,m} = \left\{
\begin{array}{ll}
U & m=0,\\
\frac{V}{m^\alpha} & m=1,\ldots,\frac{L}{2}\\
\frac{V}{\left( L-m \right)^\alpha} & m=\frac{L}{2}+1,\ldots,L-1
\end{array} \right.
\end{equation}

The non-local hopping rates $t_{j,m}$ will be also assumed to be power-law 
decaying with exponent $\beta$. We consider vanishing diagonal 
hopping ($t_{j,j}=0$) and a nearest-neighbor hopping 
\begin{equation}
t_{j,j+1}=t.
\label{def_t}
\end{equation}
Therefore, with periodic boundary conditions we have 
\begin{equation}\label{tt}
t_{0,m} = \left\{
\begin{array}{ll}
0 & m=0,\\
\frac{t}{m^\beta} & m=1,\ldots,\frac{L}{2}\\
\frac{t}{\left( L-m \right)^\beta} & m=\frac{L}{2}+1,\ldots,L-1
\end{array} \right.
\end{equation}
Since we want a finite expression for the energy per particle, we will 
consider
\begin{equation}
\alpha>1 \,,\qquad \, \beta>1.
\end{equation}
To treat the cases $\alpha \le 1$ or $\beta \le 1$ one should do a Kac 
rescaling, as it is usually done in statistical mechanics models with 
nonextensive long-range interactions \cite{ruffo09}: e.g., for $\beta \le 1$ one has 
to perform the substitution 
\begin{equation}
\label{hoppings}
t_{j,m} \to \frac{t_{j,m}}{\sum_m \frac{1}{m^\beta}}.
\end{equation}

A non-extensive ground-state energy is found in our case by 
$\alpha \to 1$ and/or $\beta \to 1$: The region 
$1 < \beta < 2$ is the weak-long-range hopping  
region. 
Recall that such region is of high interest in the study of statistical 
mechanics models with long-range couplings. 
Let us consider an Ising model of the form 
\begin{equation}
H_{\text{Ising}}=-\sum_{i,j} J_{ij} s_i s_j,  \, \,  \, \,  \, \, \left( s_i=\pm 1 \right),
\label{Ising}
\end{equation}
with power-law couplings $J_{ij} \propto 1/\mid i-j \mid^\gamma$. As usual, 
$i, j$ denote the sites of a lattice with dimension $d$. It is well known that 
for nearest-neighbor couplings (formally corresponding to 
$\gamma=\infty$) there is order at finite temperature only 
if $d \ge 2$ \cite{yeomans92}. However, if the interactions are 
sufficiently long-ranged it is possible to have 
a phase transition at finite temperature between a ferromagnetic and a 
paramagnetic phase \cite{dyson69,thouless69} even for $d=1$: If $\gamma>2$, 
then the critical temperature $T_c$ is vanishing (as in any short-range 
Ising chain \cite{yeomans92}), 
while for $\gamma \le 1$ the energy is non-extensive. After the Kac rescaling 
one easily sees that for $\gamma \le 1$ there is a phase transition having 
the critical exponents of the mean-field universality class (see e.g. 
\cite{lebellac91}). For $\gamma$ between $1$ and $2$ (weak-long-range 
region) there is a phase transition at a finite critical temperature. 
For $\gamma=2$ a Kosterlitz-Thouless transition takes place 
\cite{thouless69, anderson71, luijten01} 

Before moving on to the derivation of the Bogoliubov spectrum of elementary 
excitations, we pause here to discuss the relation between the DNLSE 
\eqref{DNLSE} and a model widely used in the treatment of long-range systems, 
i.e. the Hamiltonian mean field (HMF) model \cite{ruffo09}. We observe that 
the DNLSE Hamiltonian \eqref{def_H} can be written as the 
sum of a kinetic term $H_{\text{kin}}$ and an interaction term $H_{\text{int}}$. The kinetic 
part reads $H_{\text{kin}}=-\sum_{j,m} \psi_j^\ast t_{j,m} \psi_m $: By writing 
$\psi_j$ in terms of the local density $n_j$ and phase $\theta_j$, i.e., $\psi_j=\sqrt{n_j} 
e^{i\theta_j}$ the kinetic part reads then  
$H_{kin}=-\sum_{j,m} t_{j,m} \sqrt{n_j n_m}\cos{\left( \theta_j-\theta_m 
\right)}$. One sees than when the number fluctuations are frozen, i.e. 
$n_j \approx \rho$ (where $\rho$ is the average number of particles 
per site), 
the DNLSE Hamiltonian reduces to the 
potential energy (and it has the same equilibrium properties) of the HMF model \cite{ruffo09}
$$
H_{\text{HMF}}=- \sum_{j,m} J_{j,m} \cos{\left( \theta_j-\theta_m 
\right)},
$$
(where $J_{j,m} \equiv t_{j,m} \rho$), which is nothing 
but a long-range $XY$ model. This result is of course expected in 
the sense that 
interacting bosons on a lattice are in the $XY$ universality class, and 
interacting bosons with long-range hoppings have to be in the universality 
class of the long-range $XY$ model.

\section{Modulational stability analysis in presence of long-range interactions and hoppings} 
\label{MI_analysis}

We study in this section the Bogoliubov spectrum of elementary
excitations, describing the energy of small perturbations with
(quasi)momentum $q$ on top of a plane-wave state with
(quasi)momentum $k$ \cite{kivshar92}. The final part of the 
section is devoted to briefly recall the results of the short-range limit 
($V=0$ and $\beta \to \infty$, i.e., only nearest-neighbor hopping).
 
The stationary solutions of Eq. \eqref{DNLSE} are plane-waves
$$
\psi_j (\tau)= \psi_0  \exp{[i (k j - \nu \tau)]}:$$ 
$\nu$ is the chemical potential given (for $L \to \infty$) by
\begin{equation}\label{mu}
\hbar \nu = - 2 t \ell_\beta(k) + \rho \left[ U + 2 V \zeta 
\left(\alpha\right) \right]. 
\end{equation}
In Eq. \eqref{mu} $\rho$ is the plane-wave density 
($\rho \equiv \mid \psi_0 \mid^2$); furthermore 
$\zeta\left(\alpha\right)$ is the Riemann zeta function
\begin{equation}
\zeta\left(\alpha\right)=\sum_{m=1}^{\infty} \frac{1}{m^\alpha}
\label{zeta_R}
\end{equation}
and we introduce the function 
\begin{equation}
\ell_\beta(k)= \sum_{m=1}^{\infty} \frac{\cos{(mk)}}{m^\beta}.
\label{elle_funct}
\end{equation}
Some useful properties of the function \eqref{elle_funct} are recalled and 
discussed in the Appendix \ref{appendix_A}. 

The stability analysis of plane-waves' stationary solutions 
can be carried out by perturbing the carrier waves as
$$\psi_j (\tau) = \left[ \psi_0 + u(\tau) e^{i q j} + 
v^{\ast}(\tau) e^{-i q j} \right]
e^{i (k j - \nu t)}.$$ Retaining only the terms proportional
to $u/\psi_0$ and $v/\psi_0$, one gets
\begin{equation}\label{matrix}
i \hbar \frac{d}{d\tau}
\left(
\begin{array}{c}
u\\
v \end{array} \right)=
\left( 
\begin{array}{cc}
{\cal A} & {\cal C}\\
-{\cal C}^\ast & -{\cal B}
 \end{array} \right) \, 
\left(
\begin{array}{c}
u\\
v \end{array} \right).
\end{equation}
The quantities ${\cal A}$, ${\cal B}$, and 
${\cal C}$ in Eq. \eqref{matrix} are defined by 
\begin{equation}{\cal A}= 2 t \left[ \ell_\beta \left( k \right) - 
\ell_\beta \left( k+q \right)  \right] + \rho \tilde{V}_q,
\end{equation} 
\begin{equation}
{\cal B}= 2 t \left[ \ell_\beta \left( k \right) - 
\ell_\beta \left( k-q \right)  \right] + \rho \tilde{V}_q\end{equation} 
and 
\begin{equation}{\cal C}=\psi_0^{2} \tilde{V}_q.
\end{equation}
In the previous expressions, $\tilde{V}_q$ is the Fourier transform 
of the interaction \eqref{VV}: For finite $L$ it is 
$$
\tilde{V}_q=\sum_{m=0}^{L-1} V_{0,m} \, e^{iqm}.
$$
For $L\to \infty$ one obtains
\begin{equation}\label{V_tilde}
\tilde{V}_q=U+2V\ell_\alpha\left( q\right).
\end{equation}

From Eq. \eqref{matrix} it follows that
the excitation spectrum (i.e., the Bogoliubov dispersion relation)
for the DNLSE with power-law hoppings and interactions is
\begin{equation}
\hbar \omega_{\pm} =  \frac{{\cal A} - {\cal B}}{2} \pm 2t \sqrt{{\cal I}}, 
\label{spectrum}
\end{equation}
where $\omega_\pm=\omega_\pm(k;q)$ is a function of $k$ and $q$ (respectively
momentum of the perturbed and perturbing plane-waves). Furthermore 
\begin{align}
{\cal I}&=\frac{1}{4} \, \left\{ 2\ell_\beta(k)-\ell_\beta(k+q)-
\ell_\beta(k-q) \right\}^2+ \nonumber\\
&+\frac{\rho \tilde{V}_q}{2t} \, \left\{ 2\ell_\beta(k)-\ell_\beta(k+q)-
\ell_\beta(k-q) \right\} =\nonumber \\
&=\mathcal{F}(k;q) \left(\mathcal{F}(k;q)+\frac{\rho \tilde{V}_q}{t}\right),
\label{II}
\end{align}
where we introduced the function 
$\mathcal{F}(k;q)=\left\{ 2\ell_\beta(k)-\ell_\beta(k+q)-
\ell_\beta(k-q) \right\}/2$.
In the following we will use the convenient dimensionless parameters
\begin{equation}
\bar{U}=\frac{U \rho}{t} \,,\qquad \, \bar{V}=\frac{V \rho}{t}.
\label{UVbar}
\end{equation}
In terms of the parameters $\bar{U}$, $\bar{V}$, the quantity 
$\rho \tilde{V}_q/t$ entering Eq.~\eqref{II} reads
\begin{equation}
\frac{\rho \tilde{V}_q}{t}=\bar{U}+2 \bar{V} \ell_{\alpha}(q).
\label{notat}
\end{equation}

The carrier wave becomes modulationally unstable when
the eigenfrequencies $\omega_\pm$ in Eq. \eqref{spectrum}
acquire a finite imaginary part: The condition for stability is then ${\cal I} \ge 0$. 
A momentum $k$ is then {\em modulationally stable} if for each $q$ 
the eigenfrequencies $\omega_\pm$ are real, otherwise if it exists a $q$ 
such that ${\cal I} < 0$ then $k$ will be modulationally {\em unstable}. 
Since the eigenvalues are unaffected by substituting $k$ with $-k$ and 
$q$ with $-q$, we will consider $k$ and $q$ both belonging to the interval 
$[0,\pi]$. Notice that for $q=0$ it is ${\cal I}=\omega_\pm=0$.

When a momentum $k$ is modulationally unstable, there will be some momentum 
$q$ for which $\omega_+(k;q)$ and $\omega_-(k;q)$ have an imaginary part 
$\mathrm{Im} \, \omega_\pm (k;q)$. 
For those values of $k$ and $q$ we write 
\begin{equation}
\Gamma(k;q)=\mid \mathrm{Im} \, \omega_\pm (k;q) \mid
\label{Gamma}
\end{equation}
to quantify how unstable is the perturbed plane-wave [notice that 
the imaginary parts of $\omega_+(k;q)$ and $\omega_-(k;q)$ are by definition 
opposite in sign and equal in modulus]. 
For $k$ unstable we will use the notation 
$\Gamma_{\text{max}}(k)=\max_q {\Gamma(k;q)}$,  
where the $\max$ is taken on all the $q$ such that $\omega^2(k;q)<0$. 
The value 
of $q$ for which the maximum value of $\Gamma(k;q)$ is obtained will 
be denoted by $q_{\text{max}}$ [i.e., $\Gamma_{\text{max}}(k)=\Gamma(k;q_{\text{max}})$].

\subsection{The short-range limit}
\label{MI_short_range}

In this section we review the results and the 
region of stability for the short-range limit, having 
only local interaction ($\bar{V}=0$) and nearest-neighbor hopping: 
$t_{j,j \pm 1}=t$ and $t_{j,m}=0$ for $m \neq j\pm 1$ [formally this is the 
limit $\beta \to \infty$ in Eq. \eqref{tt}]. 

\begin{figure}[t]
\includegraphics[width=.27\textwidth,angle=270]{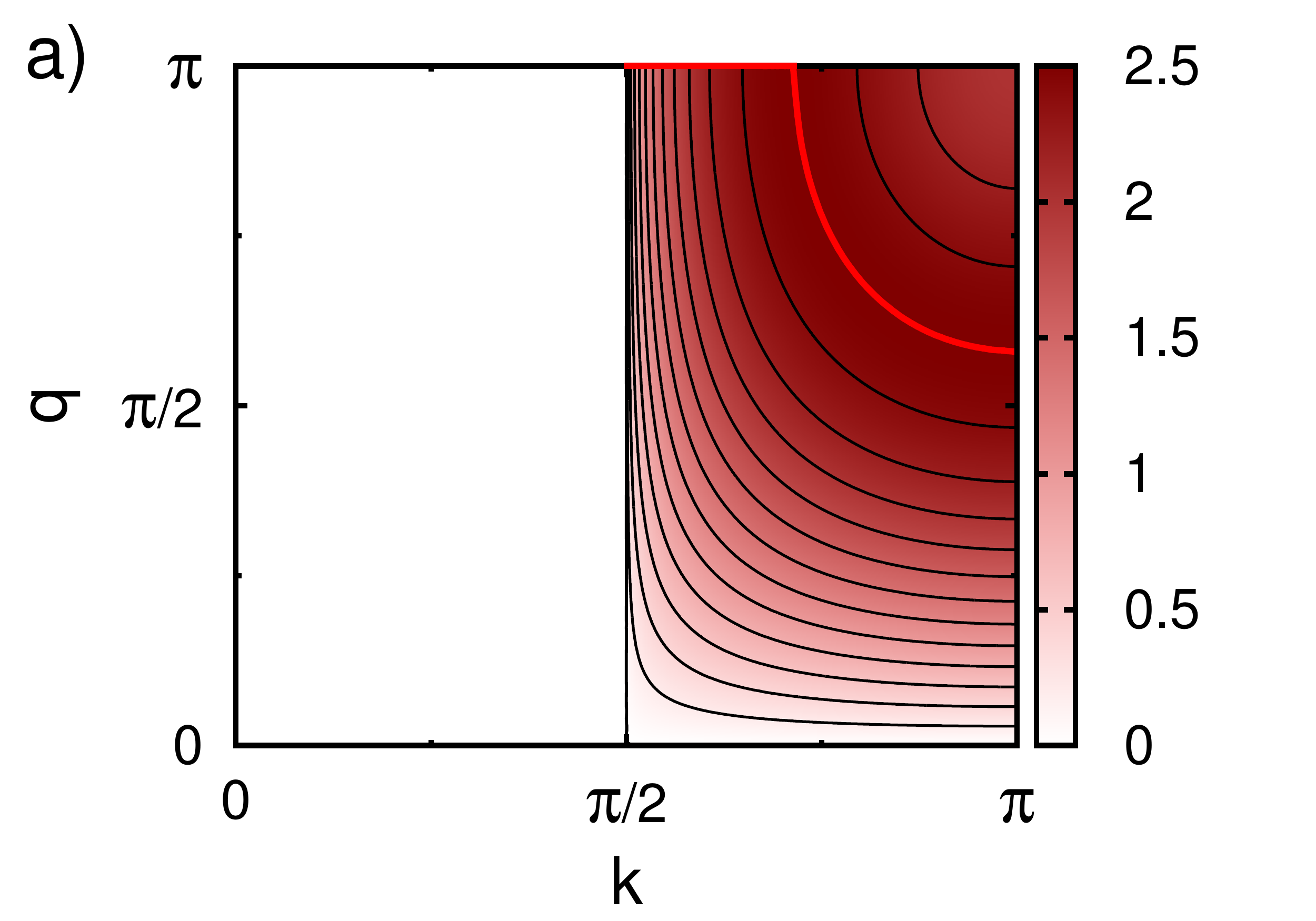}
\includegraphics[width=.27\textwidth,angle=270]{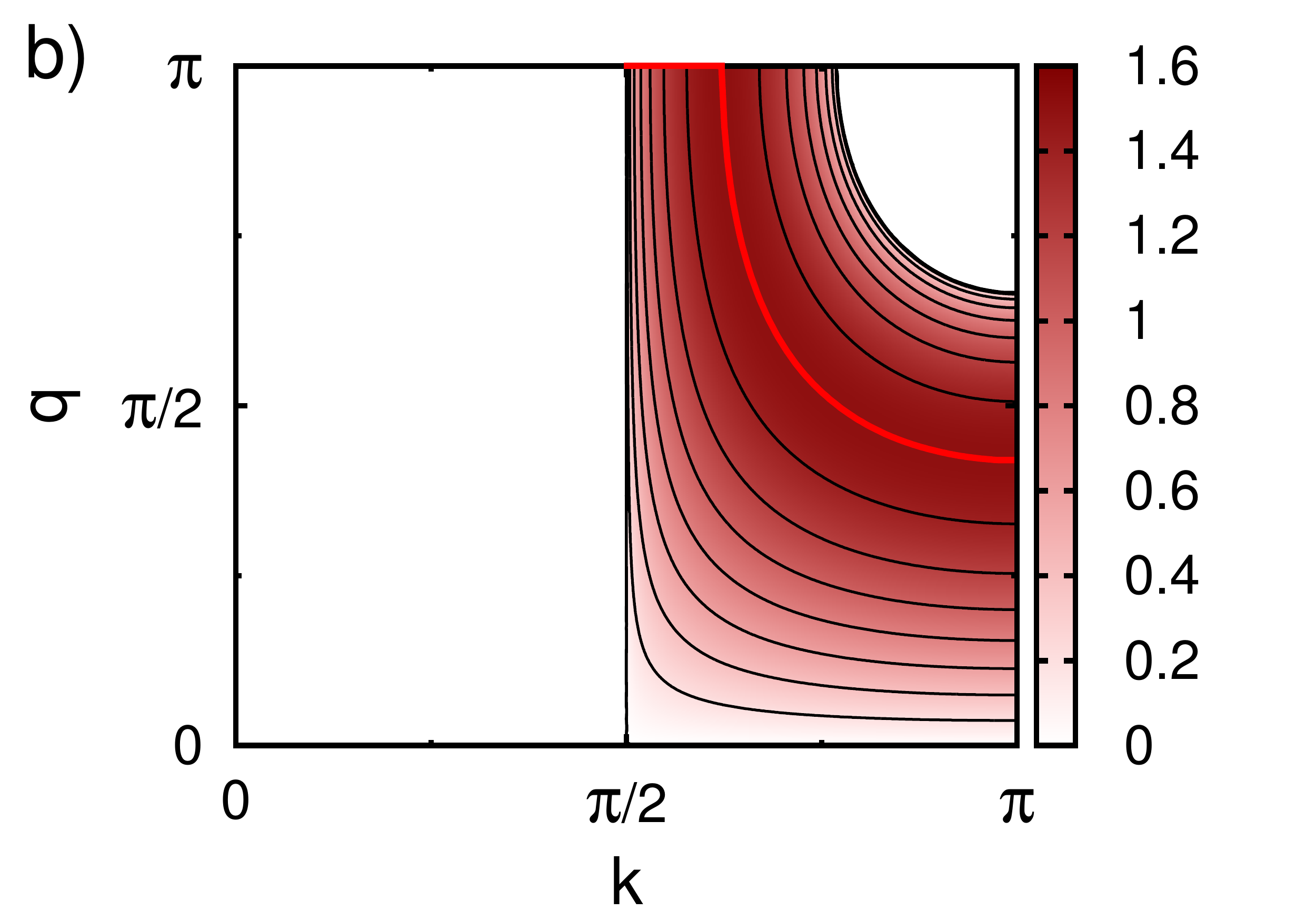}
\vspace{0.6cm}
\caption{(Color online) Stable (white) and unstable [gray, (red)] 
regions for the short-range DNLSE 
($\bar{V}=0$ and only nearest-neighbor hoppings) for $\bar{U}=2.5$ (a) 
and $\bar{U}=1.5$ (b) in the $k$ (carrier wave
momentum) $q$ (perturbing wave momentum) plane. The absolute 
value of the imaginary part of the frequencies $\Gamma(k;q)$ is depicted 
[from light to dark as $\Gamma(k;q)$ increases].
The continuous lines are equispaced isolines whose spacing is set to $0.2$.
The thick (red) line indicates the position of $q_{\text{max}}$.}
\label{fig1}
\end{figure}

It has been shown in 
\cite{kivshar92} that the onset of MI occurs at $k_{\text{cr}}=\pi/2$, i.e., 
the momenta $k<\pi/2$ are modulationally stable, while for $k>\pi/2$ are 
unstable. 
The quantity ${\cal I}$, defined in Eq.~\eqref{II} and giving the stability 
regions reads
\begin{equation}
{\cal I}=4 \cos^2{k} \sin^4{\frac{q}{2}} + 2 \bar{U} \cos{k} 
\sin^2{\frac{q}{2}}.
\label{SR}
\end{equation}
One readily sees that the momenta $k<\pi/2$ are stable. 
For $k>\pi/2$ all the momenta 
$k$ are unstable, irrespective of $U$. However, one sees that for 
$k>\pi/2$ and $\bar{U}>2$ 
each $q$ is unstable, while for $\bar{U}<2$ there are stability 
regions: These stable regions are found to be bounded by the regions 
in which $q$ is between $q_{\text{cr}}$ and $\pi$ and $k$ is between $k_{\text{cr}}$ and $\pi$, 
where
$$
k_{\text{cr}}=\pi-\arccos{\frac{\bar{U}}{2}}
$$
and 
$$
q_{\text{cr}}=2\arcsin{\sqrt{\frac{\bar{U}}{2}}}.
$$
The resulting plots of stable and unstable regions for $\bar{U}>2$ 
and $\bar{U}<2$ are in Fig.~\ref{fig1}, where we plot as well 
as contour plot the value of $\Gamma(k;q)$ in the unstable regions 
(the larger $\Gamma$, the more unstable is the dynamics). 
In Fig.~\ref{fig2} we also plot as a solid line 
the values $q_{\text{max}}(k)$ for which the maximum 
value of $\Gamma$, at the fixed $k$, is reached.

\section{Power-law interactions}
\label{MI_LR_interaction}

In this section we consider the case of power-law interactions in the presence 
of (local) nearest-neighbor hoppings ($t_{j,j \pm 1}=t$ 
and $t_{j,m}=0$ for $m \neq j\pm 1$). One has then
\begin{equation}
{\cal I}=4 \cos^2{k} \sin^4{\frac{q}{2}} + 2 \cos{k} \sin^2{\frac{q}{2}} 
\left[ \bar{U} + 2\bar{V} \ell_\alpha(q) \right].
\label{PR}
\end{equation}
We consider only the cases where $\bar V>0$ and the local interaction $\bar U$ can be 
either or positive or negative.
The cases where $\bar V$ is negative can be easily derived from the cases where
$\bar V>0$ by noticing that under the following transformation
\begin{equation}
\begin{array}{ccc}
\cos(k)& &    -\cos(k)\\
\bar U  & \rightarrow& -\bar U\\
\bar V  &                  &-\bar V \\
\end{array}
\end{equation}
one obtains the same dependence on 
$q$ and $\alpha$ of the stability
conditions that one has when $V$ is positive.
For example, if for a fixed value $\bar V= V_0 > 0$ 
and $\bar U = U_0<0$ one has that the 
momentum $\tilde k$ is stable, then 
momentum $\pi - \tilde k$ 
will also be stable for $\bar V= - V_0<0$ 
and $\bar U = - U_0>0$.

\subsection{Repulsive interactions: $U>0$, $V>0$}
\label{MI_LR_noncompeting}

\begin{figure}[t]
\includegraphics[width=.27\textwidth,angle=270]{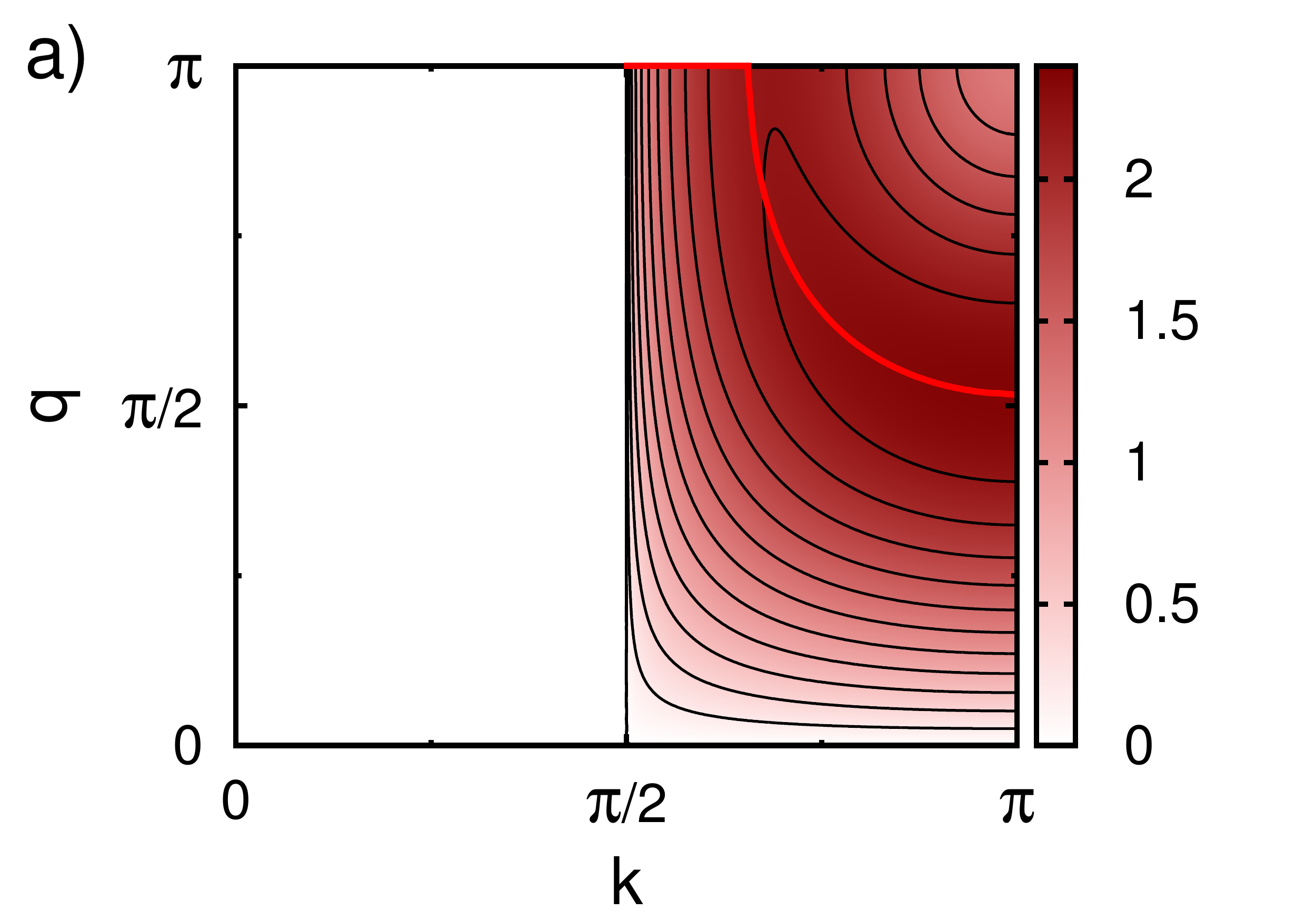}
\includegraphics[width=.27\textwidth,angle=270]{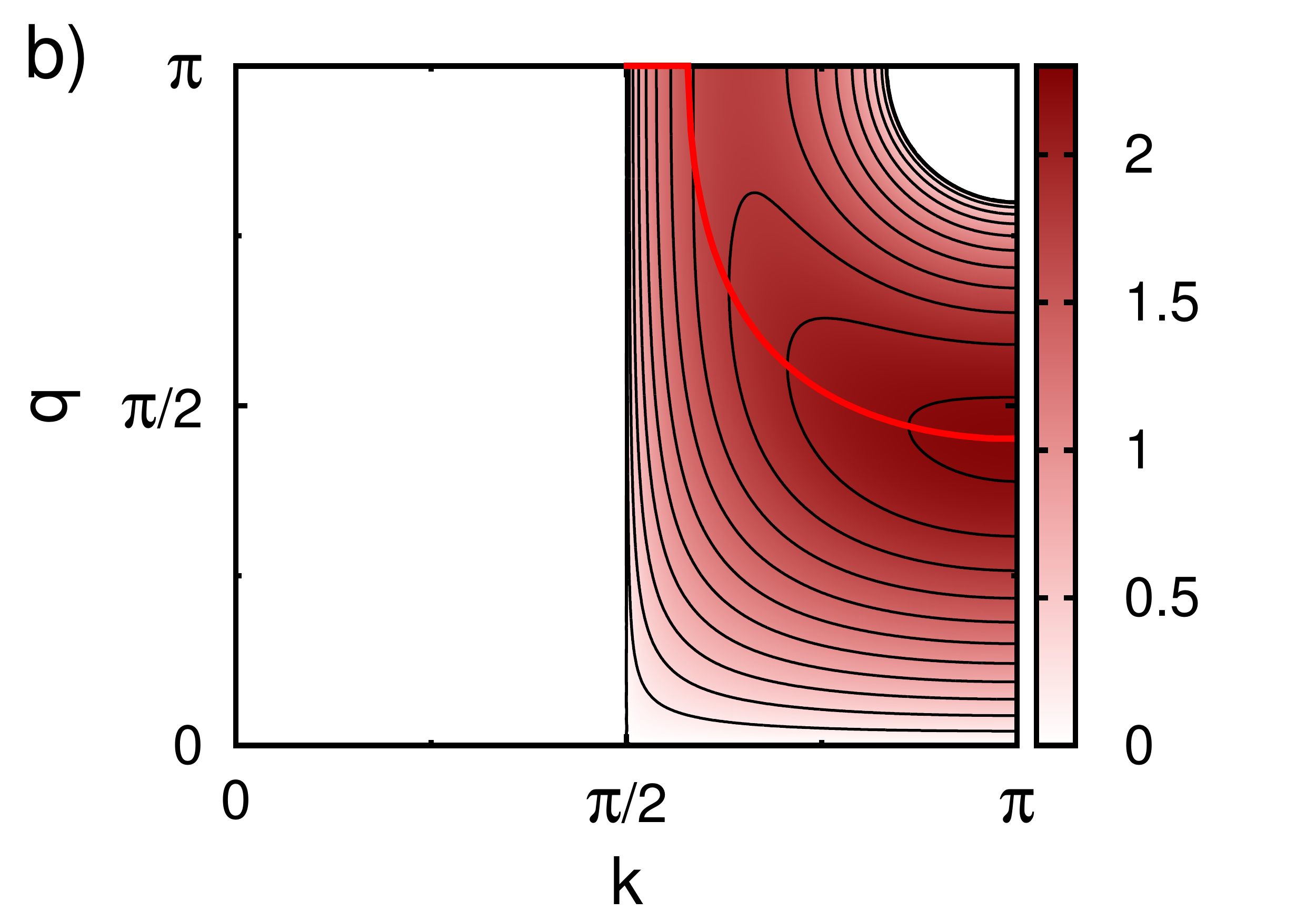}
\includegraphics[width=.27\textwidth,angle=270]{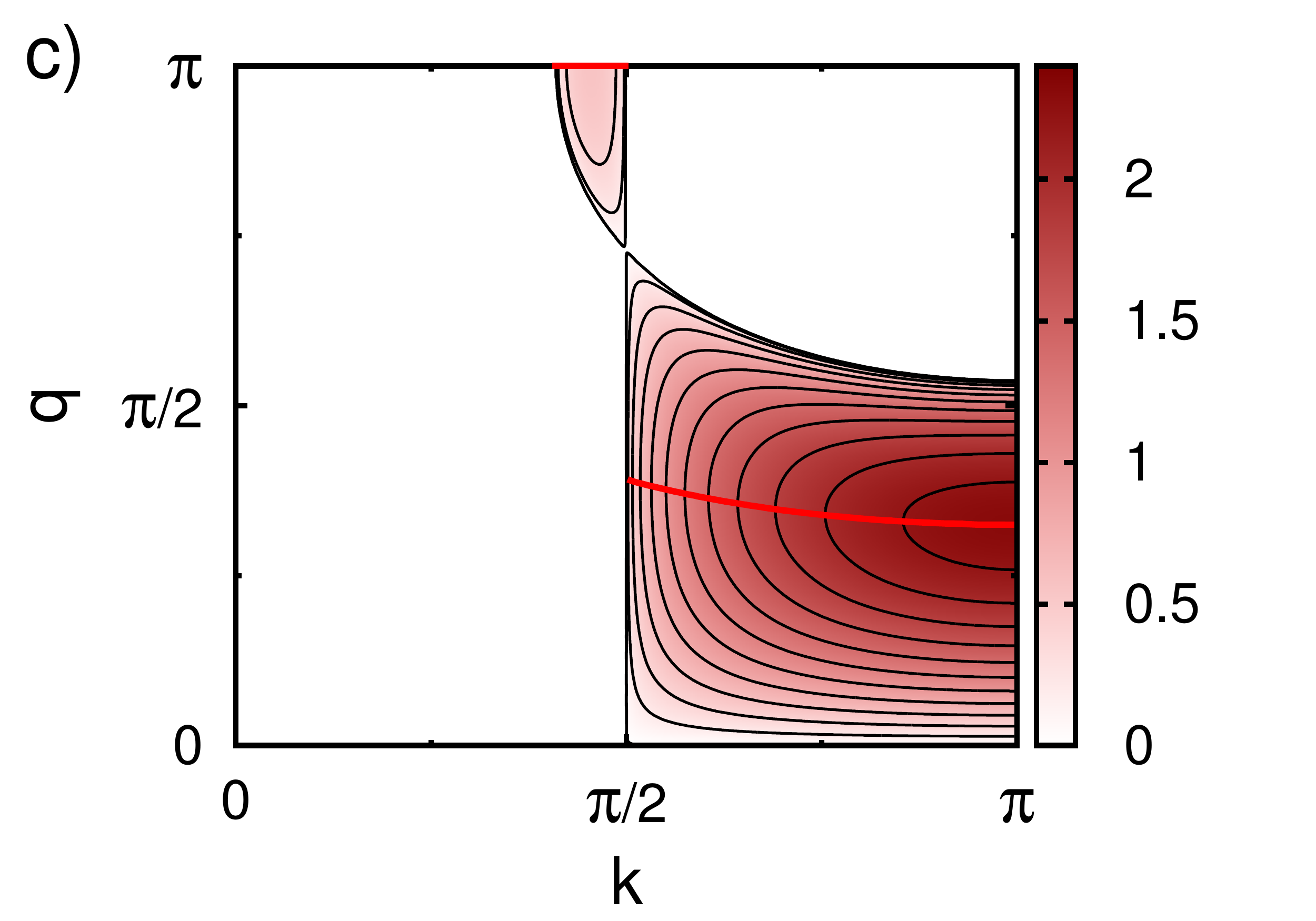}
\includegraphics[width=.27\textwidth,angle=270]{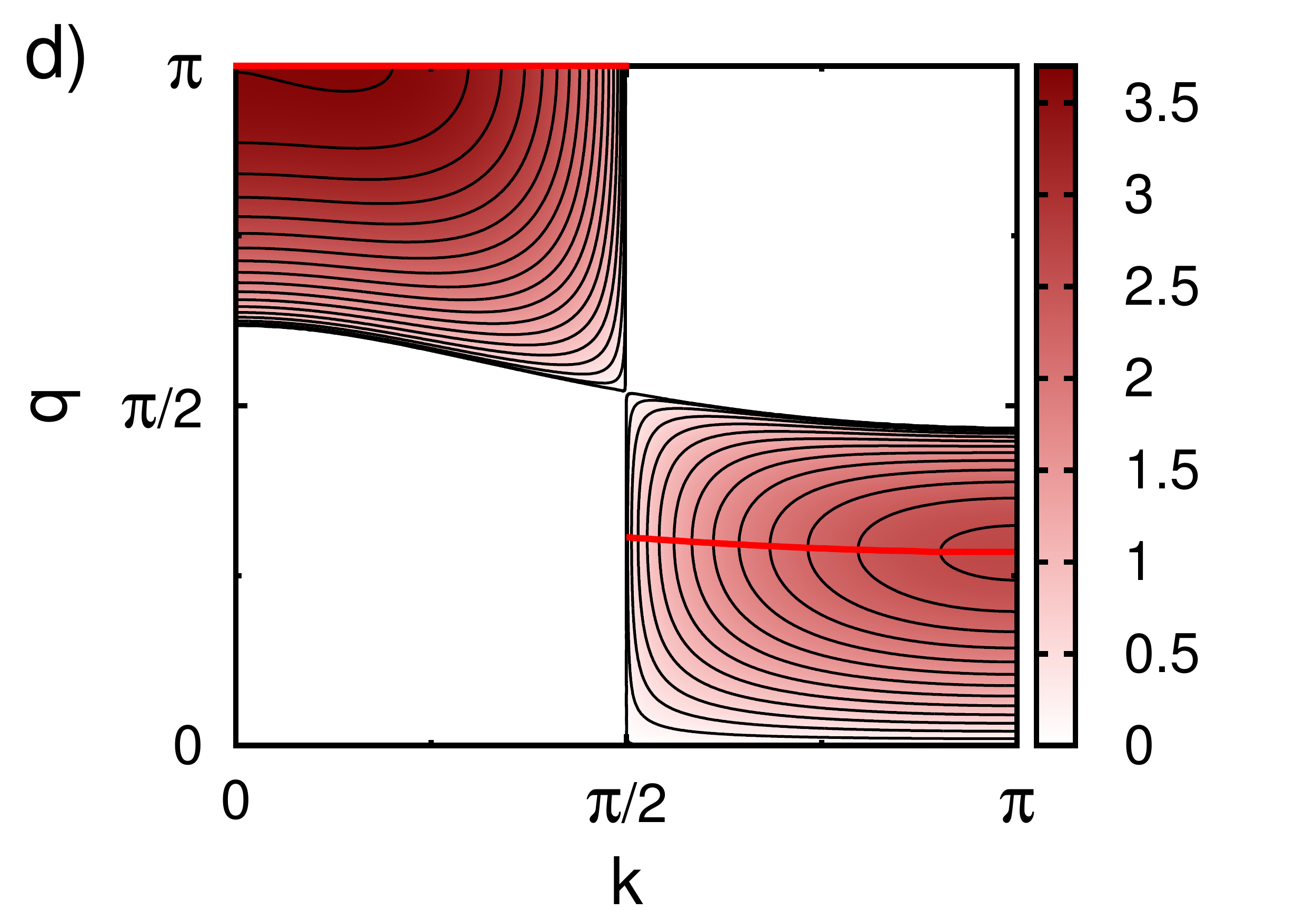}
\vspace{0.6cm}
\caption{(Color online) Stability in the $k$-$q$ plane (see caption of Fig.~\ref{fig1})
for power-law interactions in presence of local nearest-neighbor hoppings 
with $\alpha=1.5$ and $\bar{U}=2.5$. We plot $\Gamma(k;q)$ for
the four values of $\bar{V}=0.2$, $0.5$, $2$, and $4$
in panels $a$, $b$, $c$, and $d$, respectively. }
\label{fig2}
\end{figure}

We consider here $U$ and
$V$ to be positive.
Since $\partial \ell_\alpha/\partial q\le 0$ for $\alpha>1$ and 
$q \in [0,\pi]$, one can show the following:

\begin{itemize}
\item For 
$$
\bar{U} > 2\bar{V} \zeta(\alpha) (1-2^{1-\alpha})
$$
the momenta $k<\pi/2$ are stable.
\item For 
$$
\bar{U}< 2\bar{V} \zeta(\alpha) (1-2^{1-\alpha}) -2
$$
the momenta $k<\pi/2$ are unstable.
\end{itemize}
It follows that the critical value $k_{\text{cr}}$ as a function of $\bar{V}$ is 
given by 
\begin{equation}
\cos{k_{\text{cr}}}=\bar{V} \zeta(\alpha) (1-2^{1-\alpha})-\frac{\bar{U}}{2},
\label{k_cr_V}
\end{equation}
i.e., for $2\bar{V} \zeta(\alpha) (1-2^{1-\alpha})-2>\bar{U}$ one has 
$k_{\text{cr}}=0$. For $k>\pi/2$ it is easy to see that all momenta are unstable against perturbations at $q=0$. Notice that for 
$\alpha \to \infty$ (i.e., for the model having only on-site 
and nearest-neighbor interactions) Eq. \eqref{k_cr_V} 
reads $\cos{k_{\text{cr}}}=\bar{V} - \bar{U} / 2$.

The instability regions are depicted
in Fig.~\ref{fig2} as $V$ is increased - for $\bar{U}>2$, one can 
identify four regions. For $2 \bar{V} \mid \ell_{\alpha}(\pi) \mid+2<\bar{U}$ 
with $\ell_\alpha(\pi)= \zeta(\alpha) (1-2^{1-\alpha})$, 
then the momenta $k$ larger than $\pi/2$ (smaller than $\pi/2$) 
are unstable for all $q$. Increasing $\bar{V}$, one has that 
for $\bar{U}-2<2\bar{V} \mid \ell_\alpha(\pi) \mid<\bar{U}$ a stable region 
forms around $k=\pi, q=\pi$, while for  $\bar{U}<2\bar{V} 
\mid \ell_\alpha(\pi) \mid<\bar{U}+2$ an unstable region appears 
close to $q=\pi$ for momenta $k$ between $k_{\text{cr}}$ given by 
Eq. \eqref{k_cr_V} and $\pi/2$. When $\bar{U}+2>2\bar{V} 
\mid \ell_\alpha(\pi) \mid$, then all the $k$ becomes unstable and the 
instability starts from $q=\pi$. Therefore, 
the plane wave $k_{\text{cr}}$ is rendered unstable by the wave vector $q=\pi$,
thus the system is the most sensitive to short wavelength perturbations. 

\begin{figure}[t]
\includegraphics[width=.5\textwidth,angle=270]{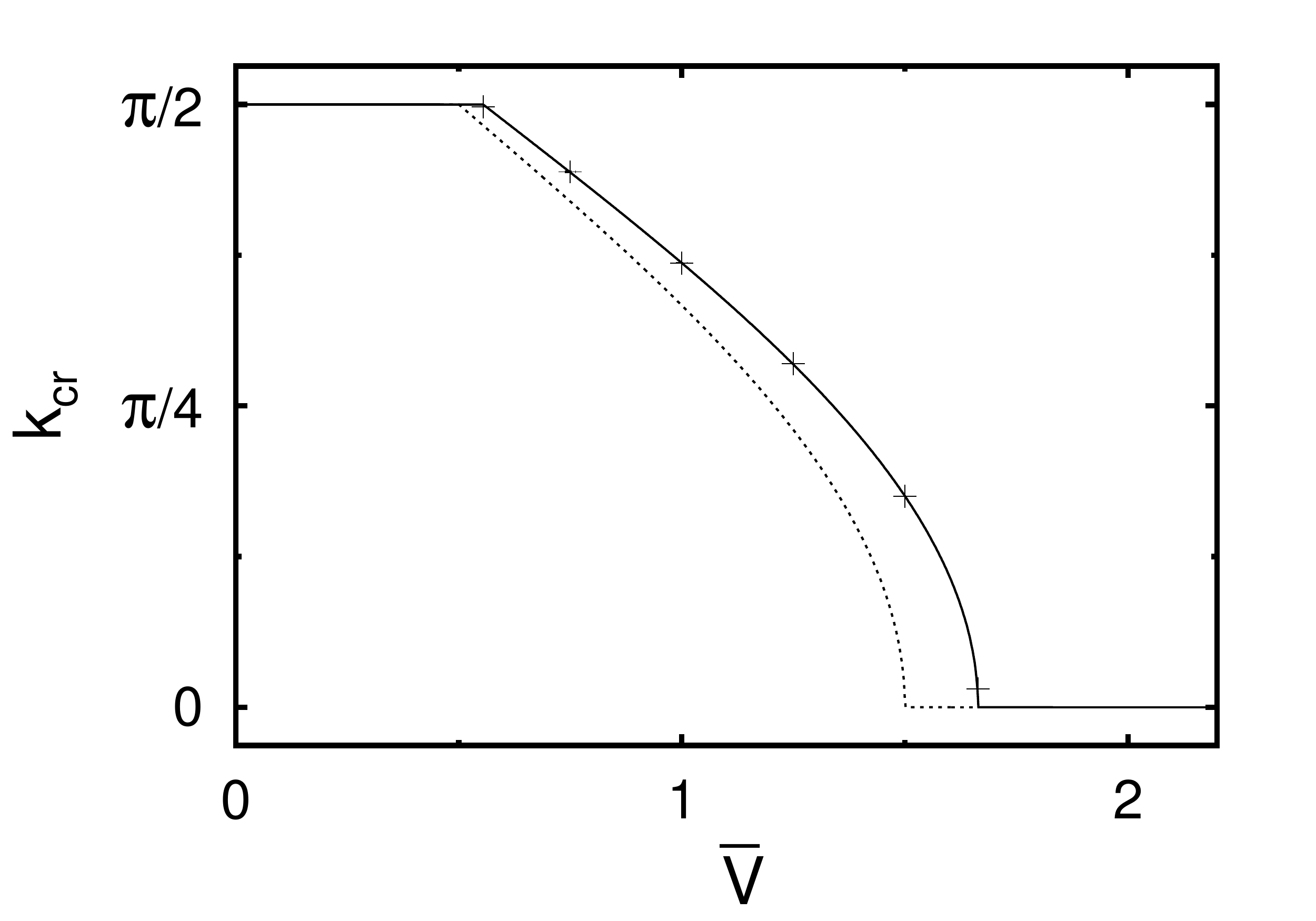}
\vspace{0.6cm}
\caption{The solid line represents $k_{\text{cr}}$ vs $\bar{V}$ from Eq.~\eqref{k_cr_V} 
(in the figure $\bar{U}=1$ and $\alpha=3$). The dots represent 
values obtained by direct simulation of the DNLSE, see the text for details 
(errors are smaller than the symbols). Notice that for $\bar{U}=1$ and 
$\alpha \to \infty$ one has $k_{\text{cr}}=\pi/2$ for $\bar{V}<0.5$ and 
$k_{\text{cr}}=0$ for $\bar{V}>1.5$. The dotted line 
is the analytical value of $k_{\text{cr}}$ vs $\bar{V}$ from Eq.~\eqref{k_cr_V} 
with $\alpha \to \infty$ and $\bar{U}=1$.}
\label{fig3}
\end{figure}

\begin{figure}[!t]
\includegraphics[width=.35\textwidth,angle=270]{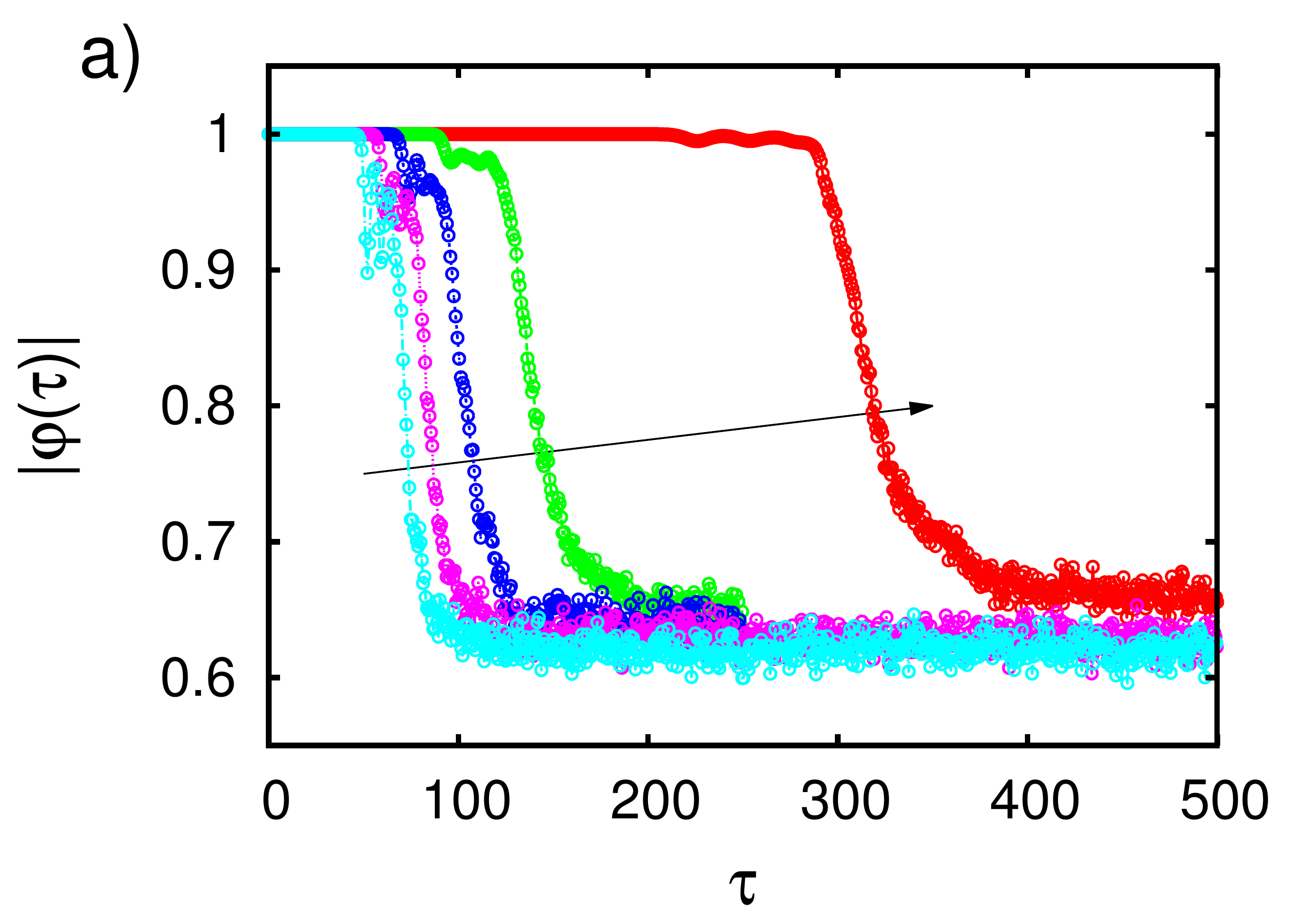}
\includegraphics[width=.35\textwidth,angle=270]{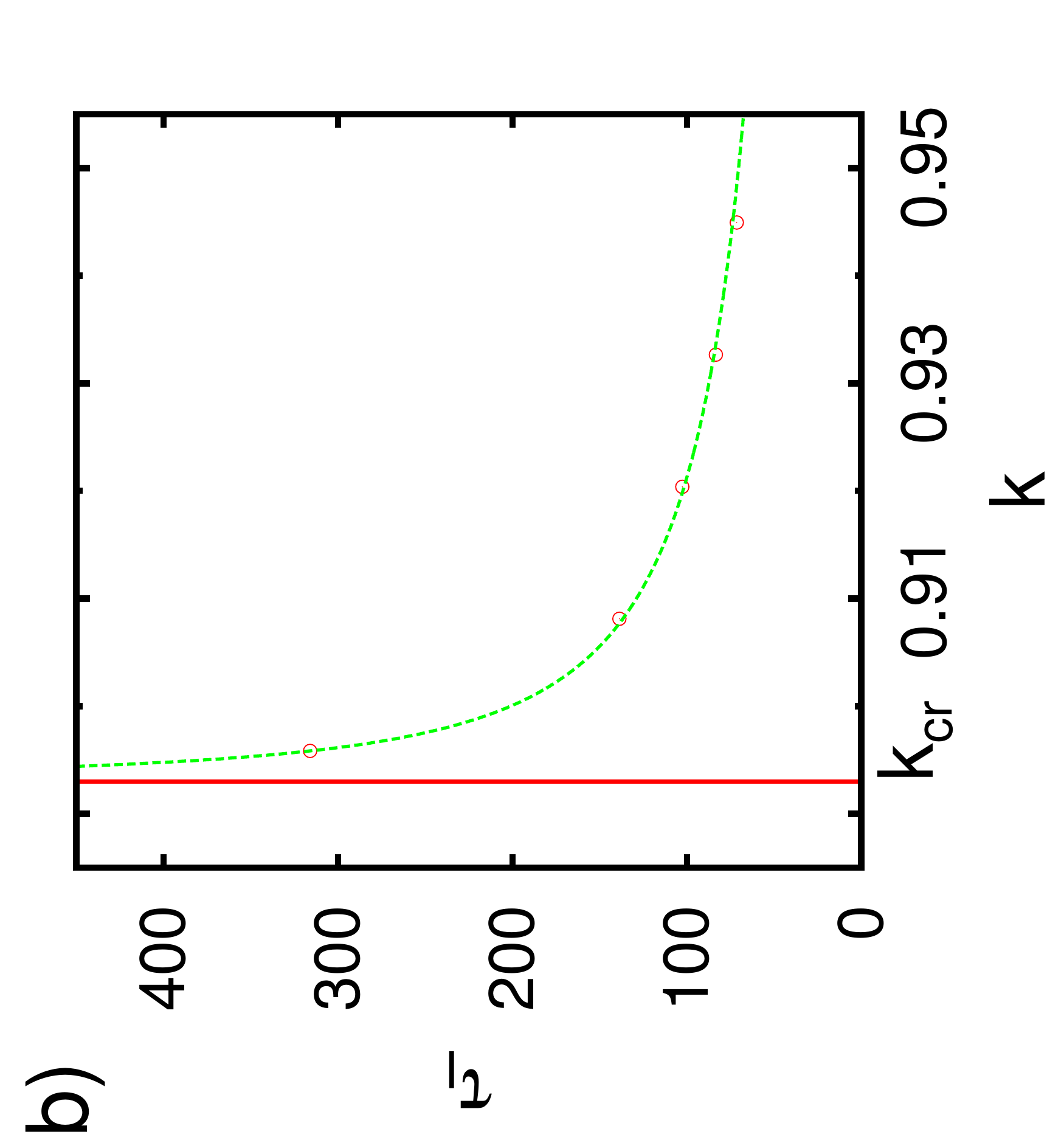}
\vspace{0.6cm}
\caption{Time evolution  (on the left) of the modulus of 
the order parameter $\varphi$, defined in Eq. ~\eqref{order_parameter}, 
for five different values 
(from left to right $k=77$, $76$, $75$, $74$, $73$, $72$ 
in units of $2\pi/L$ with $L=512$)
of carrier plane-wave wavevector for the DNLSE with parameters
$\bar{U}=1$, $\bar{V}=1.25$, and $\alpha=3$.
The lifetimes of the initial states are depicted in
the right part of the figure. These times have been fitted with a function
$\tau=\text{const} (k-k_{\text{cr}})^{-1/2}$ (dotted line) to obtain
the a numerical estimates of $k_{\text{cr}}$ (vertical line).
}
\label{fig4}
\end{figure}

The behavior of $k_{\text{cr}}$ is plotted in Fig.~\ref{fig3}. 
The analytical prediction \eqref{k_cr_V} (valid for $\alpha>1$) 
is compared for $\alpha=3$ against numerical findings obtained 
numerically solving the DNLSE finding a very good agreement [for comparison 
we also plot the analytical result \eqref{k_cr_V} for $\alpha \to \infty$]. 
The numerical solution has been obtained 
on a finite-size system (we choose $L=512$) whose coherence has been monitored
by inspecting the absolute value of the following 
order parameter \cite{smerzi02}
\begin{equation}
\varphi(\tau)=\frac{1}{L}\sum_k |\tilde{\psi}_k(\tau)|^2 e^{i k},
\label{order_parameter}
\end{equation}
where $\tilde{\psi}_k(\tau)=-\frac{1}{\sqrt{L}}\sum_m \psi_m(\tau) e^{-ikm}$
is the Fourier transform of the wave functions.
The initial wavefunction $\psi_j(\tau=0)$ is chosen as a 
plane wave with wave vector $k$  perturbed
by the highest frequency wavevector $\psi_j(\tau=0)=e^{ikj}+\epsilon 
e^{iqj}$, with $q=\pi$ (notice that $q_{max}=\pi$ for non-competing 
interactions, as one can see from Fig.~\ref{fig2}).
The ratio between the amplitudes of perturbing and perturbed wave functions
is set to be $\epsilon=10^{-3}$ with $\rho=1$.
As we can see in the left panel of Fig.~\ref{fig4}, 
if prepared in a modulationally 
unstable initial state, the system loses coherence after a time which
diverges as we approach the momentum $k_{\text{cr}}$: Denoting with $\tau_I$ the time 
after which the instability is observed using the order parameter 
\eqref{order_parameter}, and noticing that the quantity \eqref{II} 
is vanishing (at $q=\pi$) as $\sim (k-k_{\text{cr}})$, 
one can estimate $k_{\text{cr}}$ from the numerical data using 
the dependence $\tau_I \propto \left( k-k_{\text{cr}}\right)^{-1/2}$ 
(see Fig.~\ref{fig4}, right panel). 
This divergence has been used to extract the numerical
values of $k_{\text{cr}}$ shown in Fig.~\ref{fig3}.

\subsection{Competing interactions: $U<0$, $V>0$}
\label{MI_LR_competing}

For $k < \frac{\pi}{2}$ a necessary condition 
for stability is given by $|\bar U| < 2 \bar V \zeta(\alpha)$ 
that is obtained by analyzing the stability at $q=0$. The critical momentum is given by
\begin{equation} \label{kcomp}
k_{\text{cr}}= \min_{q\in [0,\pi]}\left[\arccos\left(\frac{\bar U 
+ 2\bar V \ell_\alpha(q)}{2\sin^2(\frac{q}{2})} \right)\right].
\end{equation}
all momenta $k<k_{\text{cr}}$ are stable.

An important feature of the case with competing interactions 
is that the most unstable perturbations can arise for a value 
of $q^\ast$ different from $0$ and $\pi$ (i.e., $0<q^\ast<\pi$). 
In the following we determine for what conditions at the critical 
value $k_{\text{cr}}$ there is an instability at a $q_{max}=q^\ast \neq 0, \pi$. 
We will 
refer to these values of $q^\ast$ as to {\em finite} values 
for the occurrence of modulational instability: This is because 
the instability develops on a length scale $\sim 1/q^\ast$. It is intended 
that if the instability arises at $q=\pi$, this develops on a length scale 
of the order the lattice unit, while an instability at $q=0$ involves 
length of the order of the lattice size: The later of these will be the case 
for non-local hoppings with weak-long-range exponents $1<\beta<2$. 

\begin{figure}[t]
\includegraphics[width=.55\textwidth,angle=270]{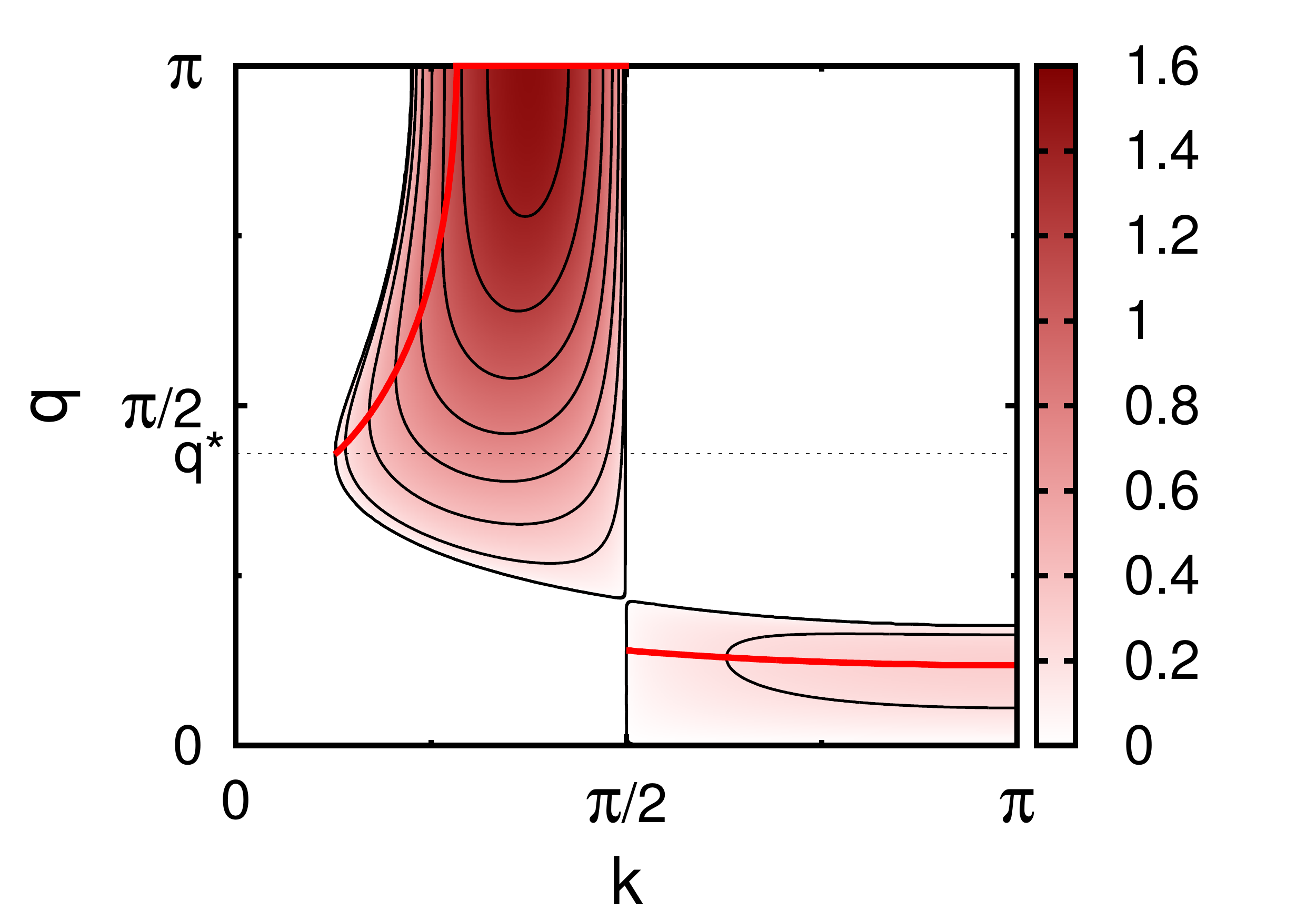}
\vspace{0.6cm}
\caption{(Color online) Stability regions (see caption of Fig.~\ref{fig1}) for the case of
competing interactions with $\alpha=2$, $\bar{U}=-0.7$ and $\bar V=0.5$.
The value of the most unstable $q$ which defines $k_{\text{cr}}$ is denoted 
by $q^*$.}
\label{fig5}
\end{figure}

An example of a finite $q^\ast$ is shown in Fig.~\ref{fig5}. 
To see how this can arise we analyze for simplicity the situation 
at $k=0$, which 
can be easily generalized to a finite value in the interval 
$0<k<\pi / 2$.
To observe an instability region like the one in Fig.~\ref{fig5} 
we have to impose that the equation  
\begin{equation}
h(q) = |\bar U|
\label{equaz}
\end{equation}
with
\begin{equation} 
h(q) = 2 \sin\left(\frac{q}{2}\right)^2 + 2\bar V \ell_\alpha(q),
\end{equation} 
obtained from Eq. \eqref{PR} by the substitution $k=0$,
has exactly one finite solution for fixed values
of the parameters $(\alpha,\bar U, \bar V)$ and no solutions 
for smaller values of $\bar U$. 
After some algebra one can derive a set of conditions on the coefficients
$(\alpha, \bar V)$ such that the curve $h(q)$ possesses one minimum.
It turns out that, if $\bar U = \min_{q\in (0,\pi)}h(q)$, then there is a
$q^*\in \left(0,\pi \right)$ such that the instability region is
tangent to the $k=0$ axis. For smaller values of $\bar U$ the previous
equation $h(q)=|\bar U|$ does not have any solution and one obtains
a finite value of $k_{\text{cr}}$ given by Eq. \eqref{kcomp}. 

We can thus assert that the presence of competing interactions
may originate the wavelength $q^*$ smaller than $\pi$ (and larger than $\pi$), 
unlike the case of the noncompeting interaction examined 
in the previous section for which the system is most sensible to 
perturbations at $q=\pi$.
This is a general feature of systems with 
competing interactions acting on different scales 
which, if properly tuned, give
rise to the birth of a new intermediate lengthscale (in this case
of the order of $1/q^*$). For similar phenomena
ultimately leading to stripe formation and more generally 
spatially modulated patterns in different contexts see, 
e.g., \cite{seul95, chakrabarty11}.

For completeness we list the conditions on the parameters $(\alpha, \bar V)$
such that the function $h(q)$ has a minimum for a finite value 
$q^\ast$ of the perturbing wavevector:
\begin{itemize}
\item for $1<\alpha\le \alpha^*$
\begin{equation}
\bar V < \frac{1}{(2-2^{1-\alpha})\zeta(\alpha)}
\label{tom1}
\end{equation}
\item for $\alpha^*<\alpha\le 3$
\begin{equation}
\bar V < \frac{1}{2(1-2^{3-\alpha})\zeta(\alpha-2)}
\label{tom2}
\end{equation}
\item for $\alpha > 3$
\begin{equation}
\frac{1}{2|\zeta(\alpha-2)|}<\bar V < \frac{1}{2(1-2^{3-\alpha})\zeta(\alpha-2)}.
\label{tom3}
\end{equation}
\end{itemize}
(similar results are found for $0<k<\pi/2$). 
The value $\alpha^*$ in Eq.~\eqref{tom2} is given 
by $\approx 1.513$, the unique solution of the equation:
\begin{equation}
2(1-2^{3-\alpha})\zeta(\alpha-2) = (1-2^{1-\alpha})\zeta(\alpha).
\end{equation}

The condition ensuring the stability for every $k \in [\frac{\pi}{2},\pi]$
is given by:
\begin{equation}
|\bar U| > 2 \bar V \zeta(\alpha).
\end{equation}

\begin{figure}[t]
\includegraphics[width=.27\textwidth,angle=270]{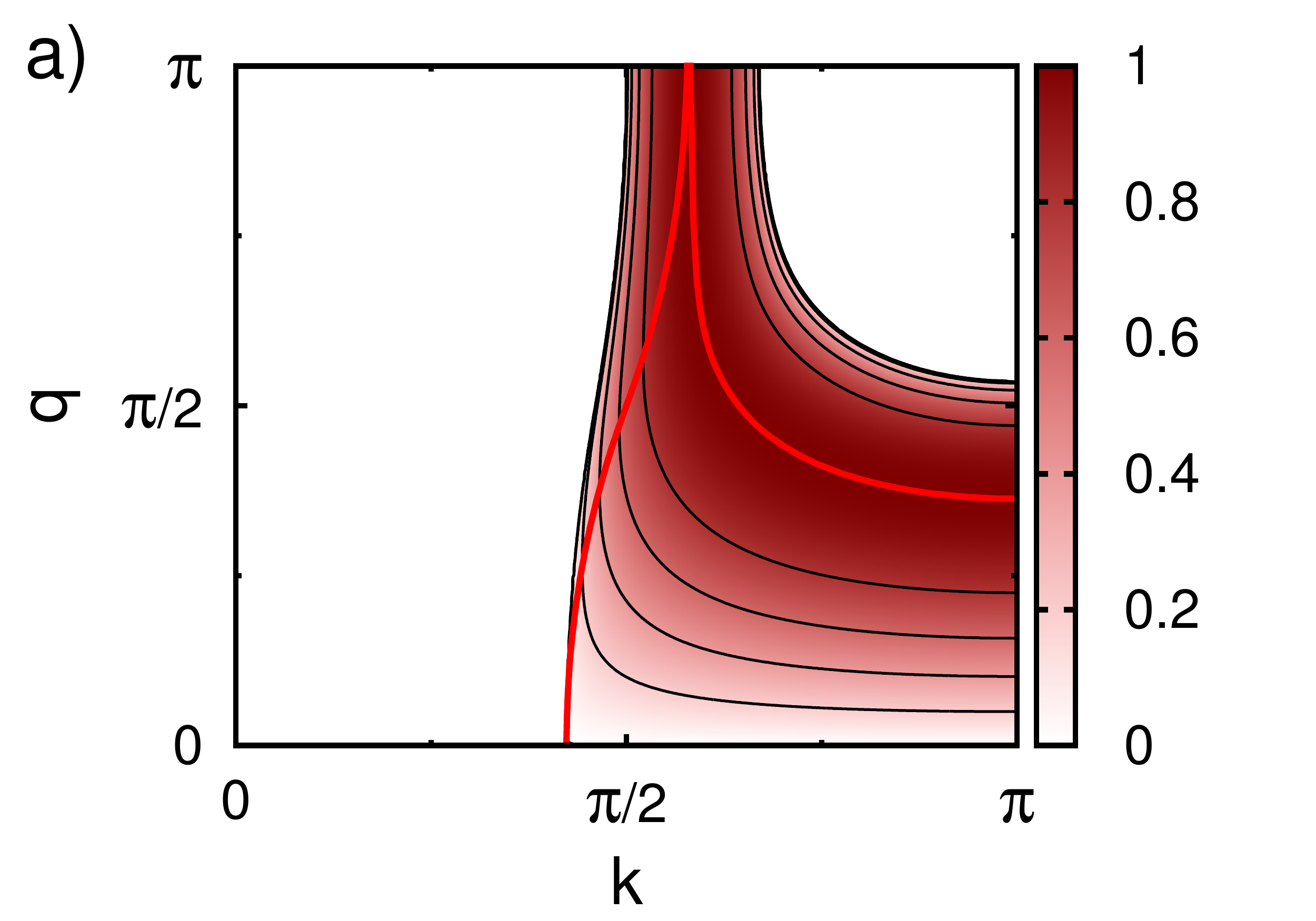}
\includegraphics[width=.27\textwidth,angle=270]{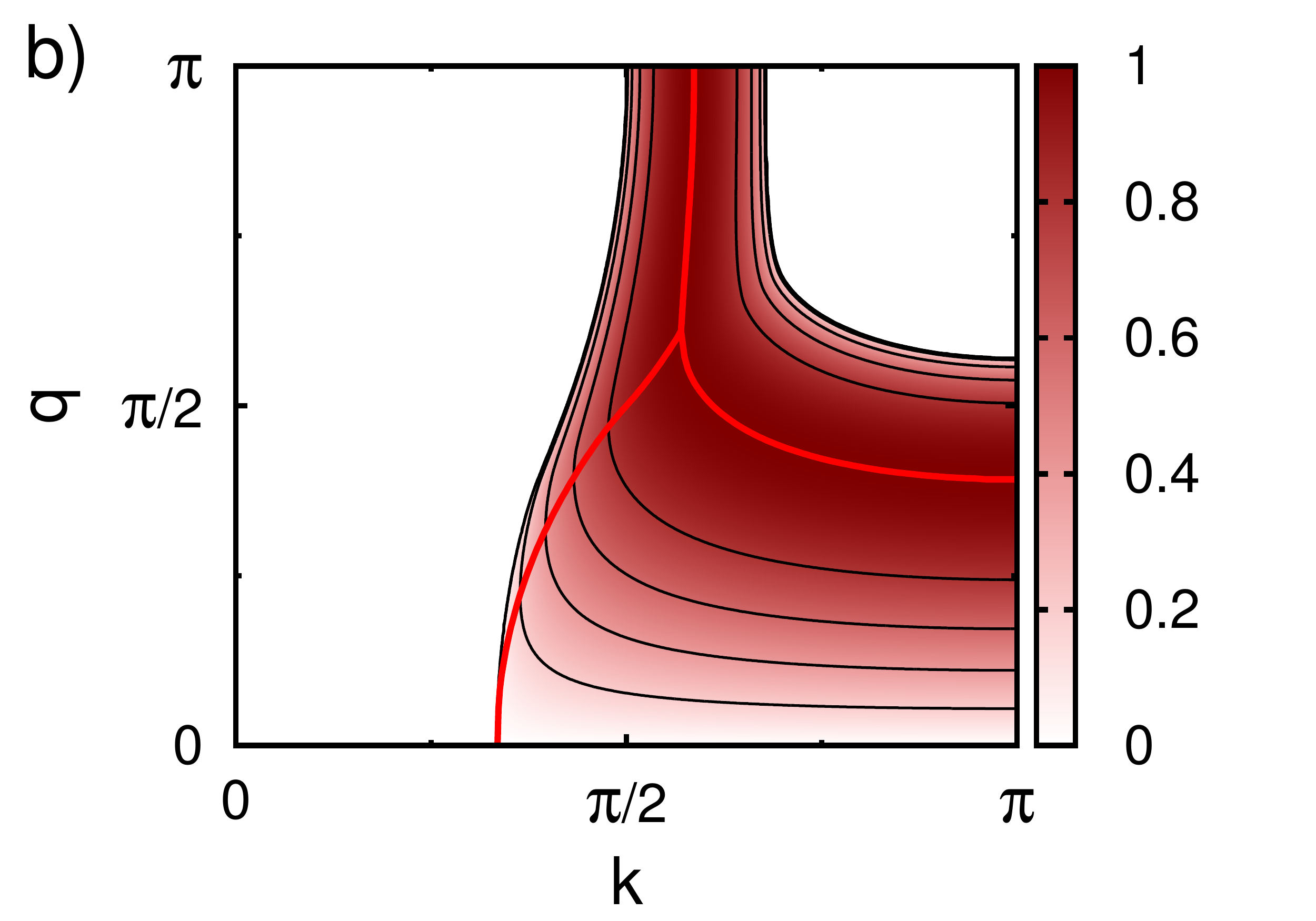}
\includegraphics[width=.27\textwidth,angle=270]{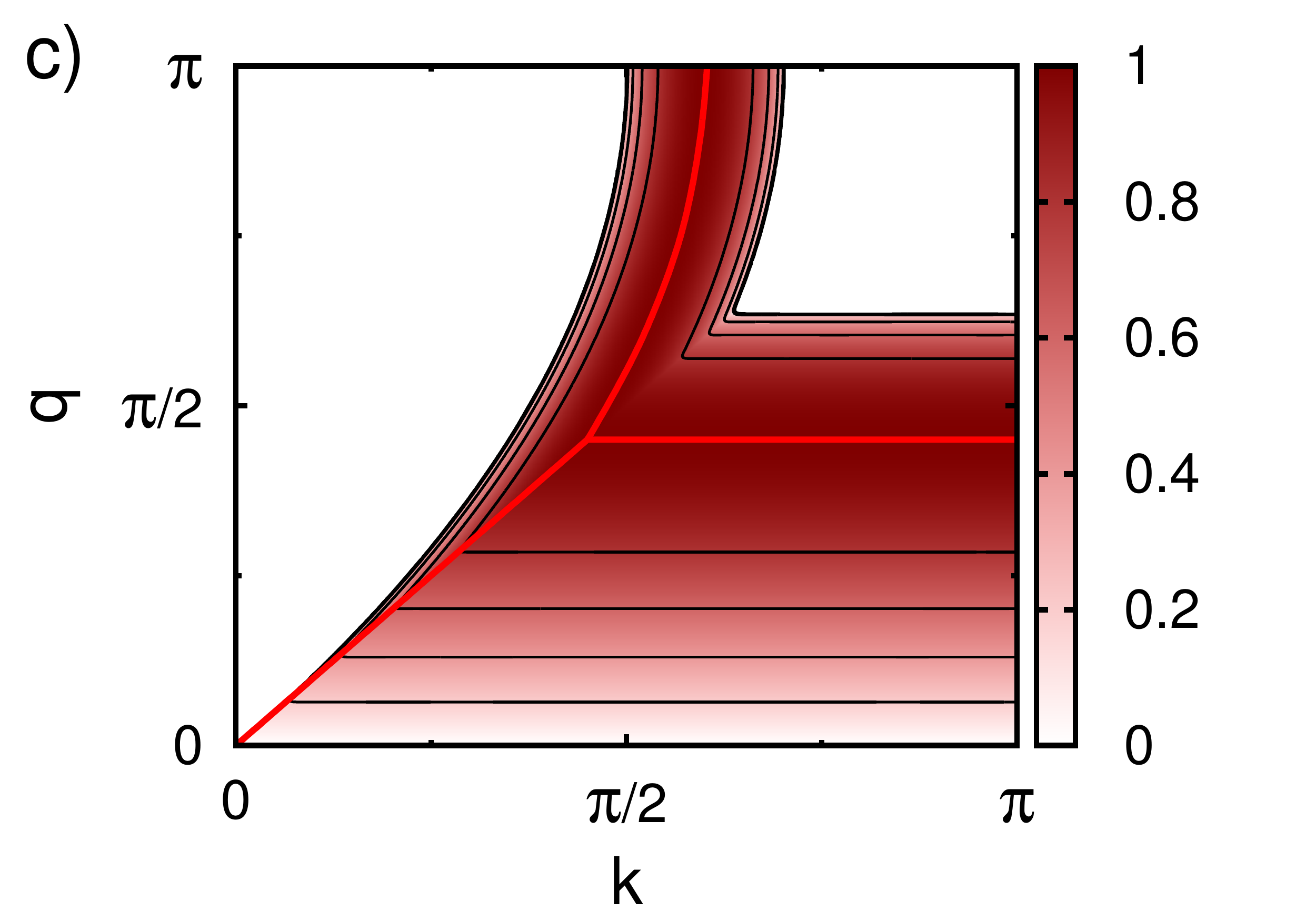}
\includegraphics[width=.27\textwidth,angle=270]{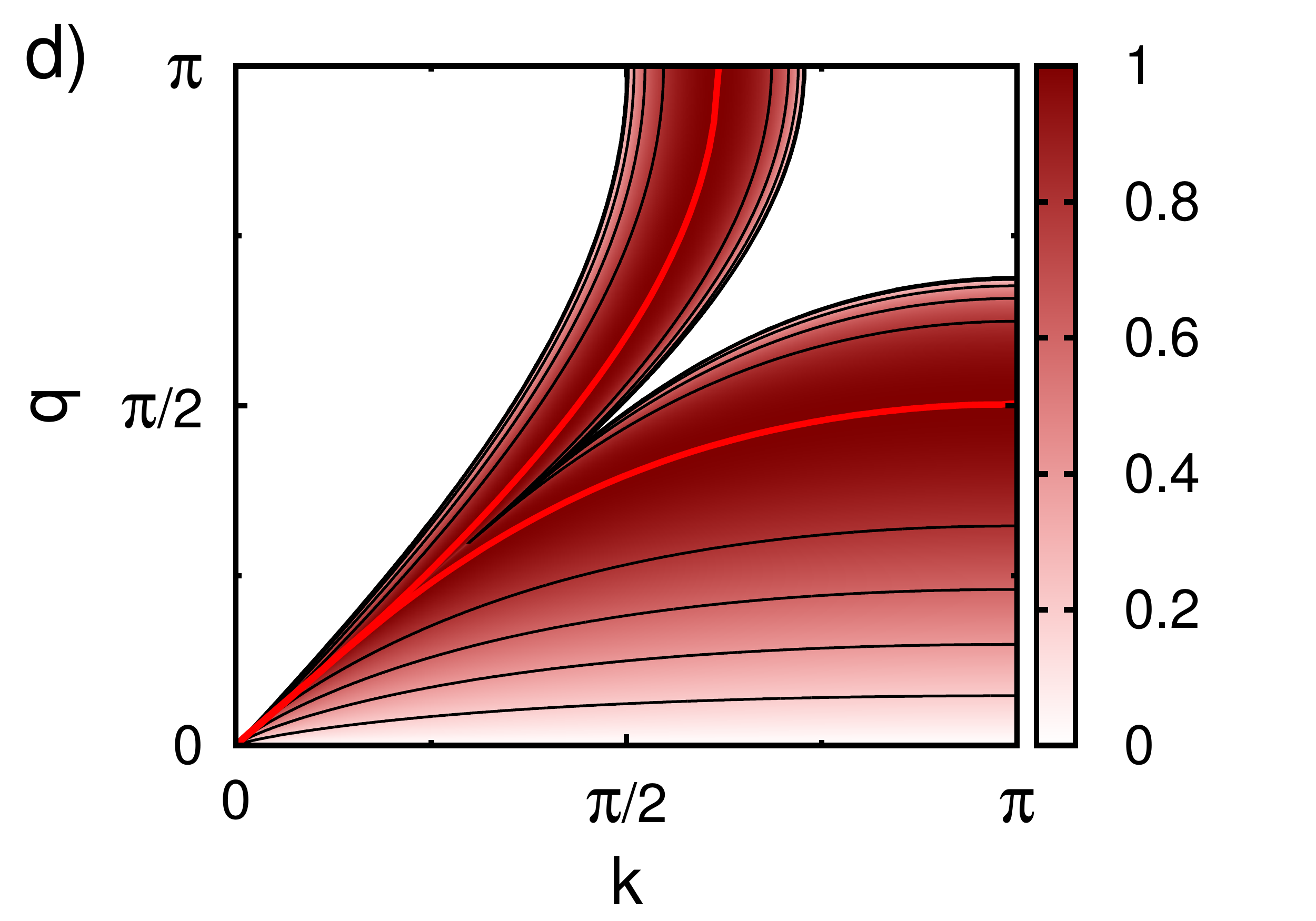}
\vspace{0.6cm}
\caption{(Color online) Stability regions (see caption of Fig.~\ref{fig1}) for the case of
long-range hopping and local interaction $\bar{U}=1$ and $\bar V=0$.
Panels (a), (b), (c), and (d) refer to the values of
$\beta=4$, $3$, $2$, $1.5$.}
\label{fig6}
\end{figure}

\section{Power-law hoppings}
\label{MI_LR_hopping}

\begin{figure}[t]
\includegraphics[width=.5\textwidth,angle=270]{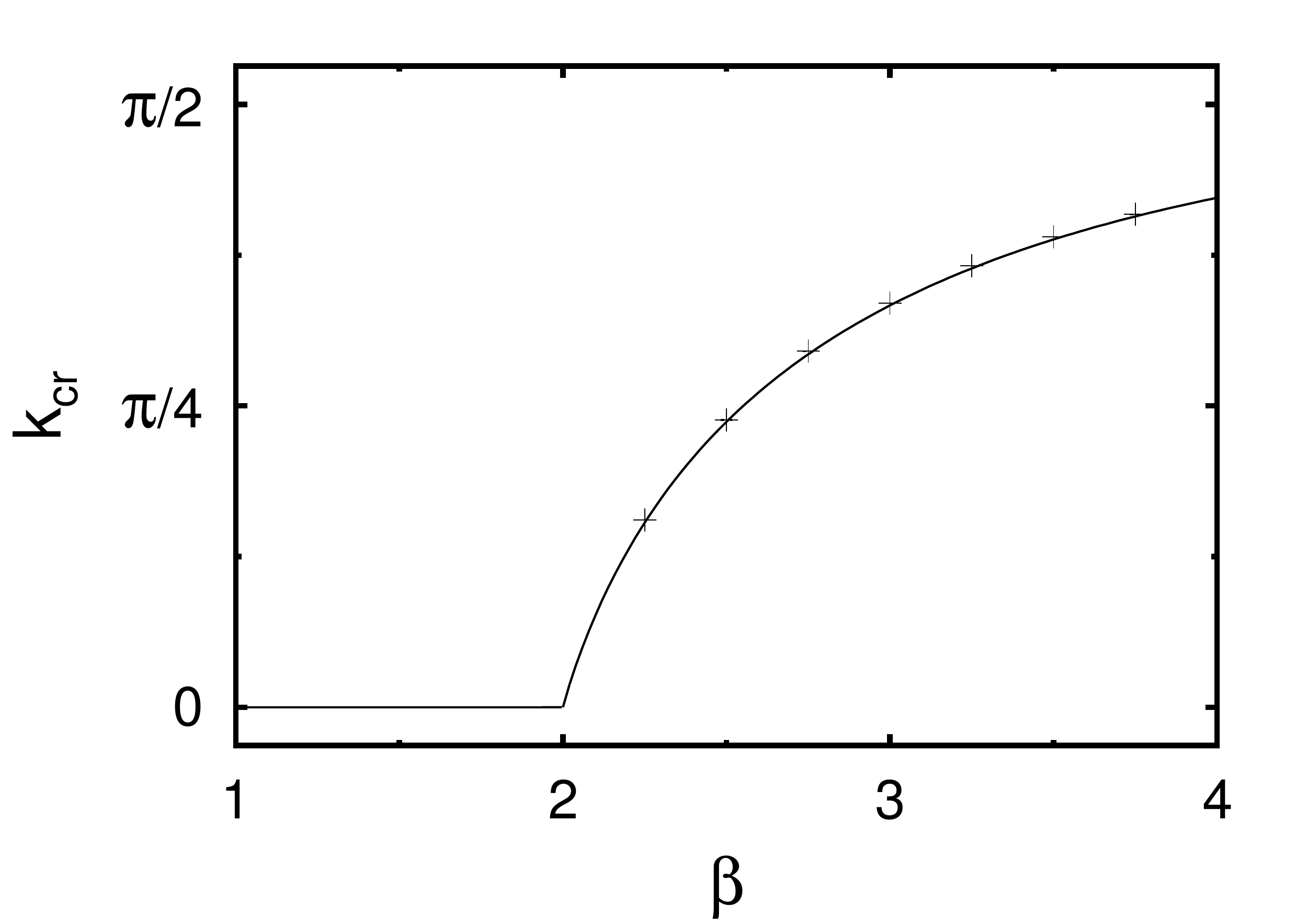}
\vspace{0.6cm}
\caption{Critical momentum $k_{\text{cr}}$ vs the power-law exponent $\beta$ of 
power-law hoppings with $\bar{V}=0$. Crosses are numerical data.}
\label{fig7}
\end{figure}

In this section we consider the case of power-law 
nearest-neighbor hoppings, with exponent $\beta>1$. One has
\begin{align}
{\cal I}&=\frac{1}{4} \, \left\{ 2\ell_\beta(k)-\ell_\beta(k+q)-
\ell_\beta(k-q) \right\}^2+ \nonumber\\
&+ \frac{\bar{U}}{2} 
 \left\{ 2\ell_\beta(k)-\ell_\beta(k+q)-
\ell_\beta(k-q) \right\}=\nonumber\\
&=\mathcal{F}(k;q) (\mathcal{F}(k;q)+\bar{U})\label{I_LR_hop}
\end{align}
While in the cases considered in Sec.~\ref{MI_LR_interaction} 
the instabilities arise in the higher frequency range of $q$ [$q=\pi$] for 
non-competing interaction or at a finite value of $q$ 
[$q=q^\ast \in (0, \pi/2)$] for competing interactions, 
with non-local hoppings even the long-wavelength perturbations
can affect significantly the stability properties of the system.
This can be verified by an inspection of the behavior of 
${\cal I}$ for small $q$. One finds for $q \to 0$
$$
{\cal I} \approx -\frac{\bar{U}q^2}{4} \frac{\partial^2 \ell_\beta}
{\partial k^2}.
$$
The investigation of the behavior of the second derivative of $\ell_{\beta}(k)$ 
reveals that $\partial^2 \ell_\beta /\partial k^2$ is positive for $1<\beta<2$ for each $k$. 
It follows that
\begin{equation}
k_{\text{cr}}=0 \, \, \, \text{for} \, \, \, 1<\beta<2,
\end{equation}
i.e., the modulational stability regions shrink to zero in the weak-long-range 
regime, irrespective of $\bar{U}$ and $\bar{V}$ (we assume for simplicity 
in this section $\bar{U}>0$ and $\bar{V}>0$). The same analysis shows that 
for $\bar{V}=0$ the critical value $k_{\text{cr}}$ does not depend on the specific 
value of $\bar{U}$ for $\beta>2$.
The above scenario is confirmed by Fig.~\ref{fig6} where we
plot $\Gamma(k;q)$ as $\beta$ is increased.
The values of $k_{\text{cr}}$ as a function of $\beta$ are 
plotted in Fig.~\ref{fig7}. Notice that $k_{\text{cr}}$ tends to $\pi/2$ when the
hopping exponent approaches the short range limit $\beta \to 
\infty$. As was done in Sec.~\ref{MI_LR_interaction} we compare the analytical
results with numerical simulations of the DNLSE obtaining a very 
good agreement.

It should be stressed that the lifetimes of the modulationally unstable
states in this case is typically longer than the ones encountered in 
Sec.~\ref{MI_LR_interaction}. This will be supported in the next
section where we directly compare instabilities arising from
the long-range interaction and hopping, respectively.

When we introduce the long-range interaction $\bar V\neq 0$ the instability
at $q=0$ due to long-range hopping remains unchanged since it is
due to the vanishing of $\mathcal{F}(k;q)$ 
[see Eq.~\eqref{I_LR_hop}], while the instabilities already discussed 
in Sec. \ref{MI_LR_interaction} are possibly generated. More precisely, 
an instability region at a finite value $q=q^*$ ($q^*=\pi$ for 
non-competing interactions) arises in the presence of $V$, however, 
for $V$ much smaller than $U$ (and smaller than a critical value $V_c$) 
then $k_{\text{cr}}$ is determined only by the $q=0$ 
instability driven by non-local hoppings and it is again given by 
the value at $V=0$, as shown in Fig.~\ref{fig7}.  When $V>V_c$ 
the contribution of both instabilities
has to be taken into account to determine $k_{\text{cr}}$, 
with $q=0$ instabilities 
generated by long-range hopping and $q=\pi$ 
(or $q=q^*<\pi$ if the competition is 
sufficiently strong) instabilities due to the interaction.
We comment on these scenarios in the next section.

\section{Comparison of instabilities arising with non-local interactions and hoppings}
\label{MI_interaction_hopping_comparison}

The hopping instability described in Sec.~\ref{MI_LR_hopping}
exhibits an important difference with the one arising from the interaction 
described in Sec.~\ref{MI_LR_interaction} since 
$k_{\text{cr}}$ becomes unstable for $q=0$ perturbations, i.e.,
with a size of the order of the system's size.

\begin{figure}[t]
\includegraphics[width=.23\textwidth,angle=270]{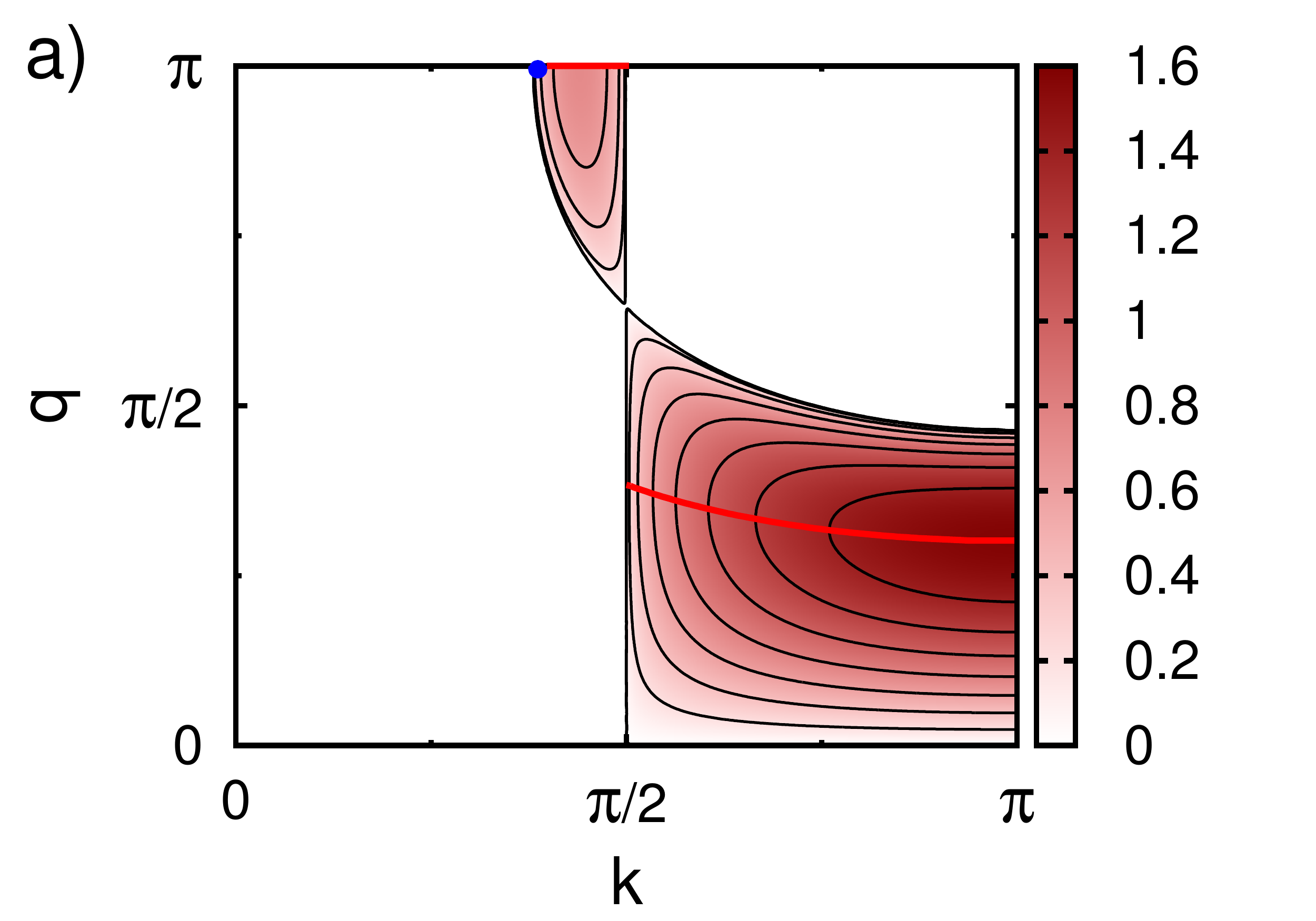}
\includegraphics[width=.23\textwidth,angle=270]{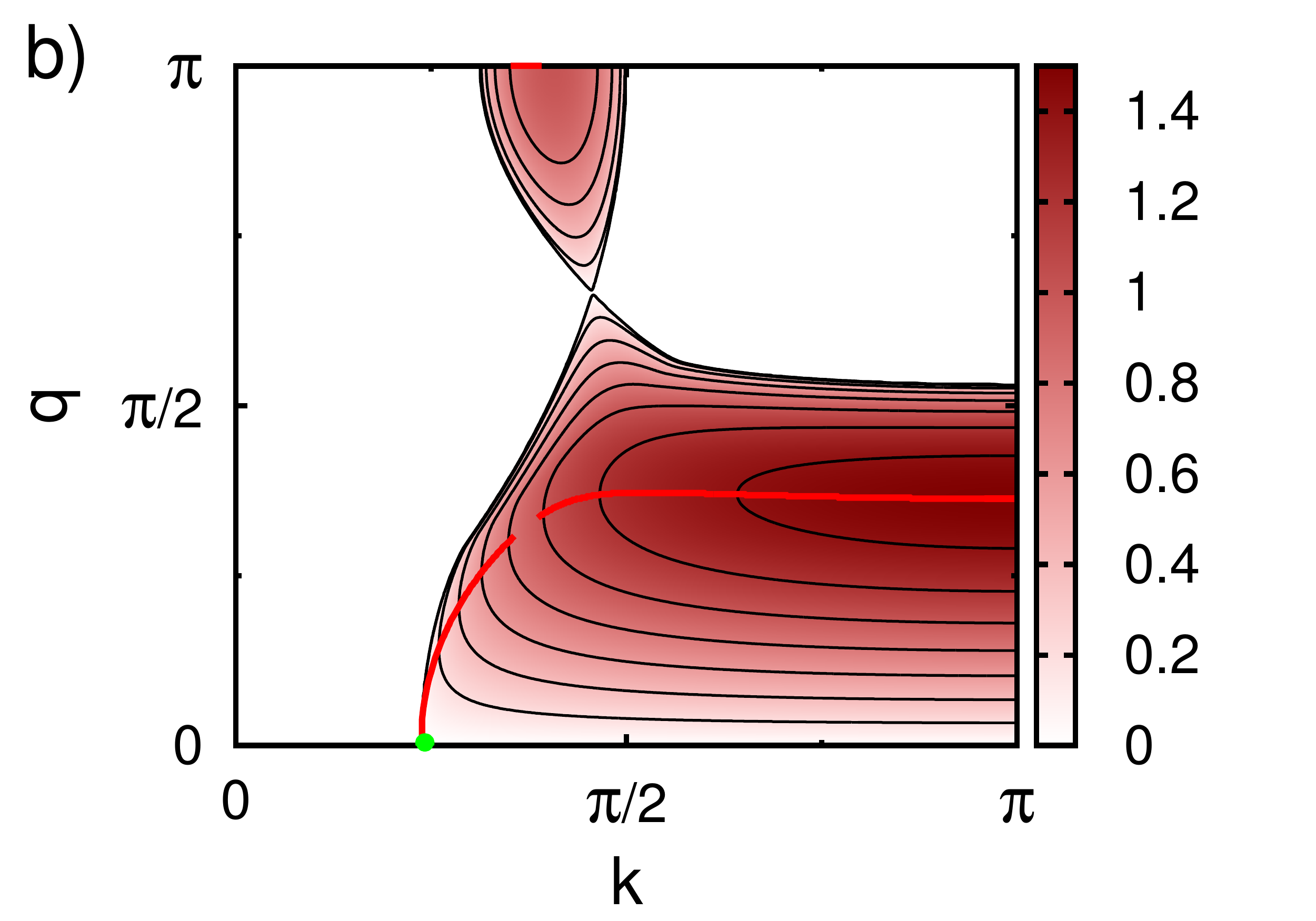}
\includegraphics[width=.23\textwidth,angle=270]{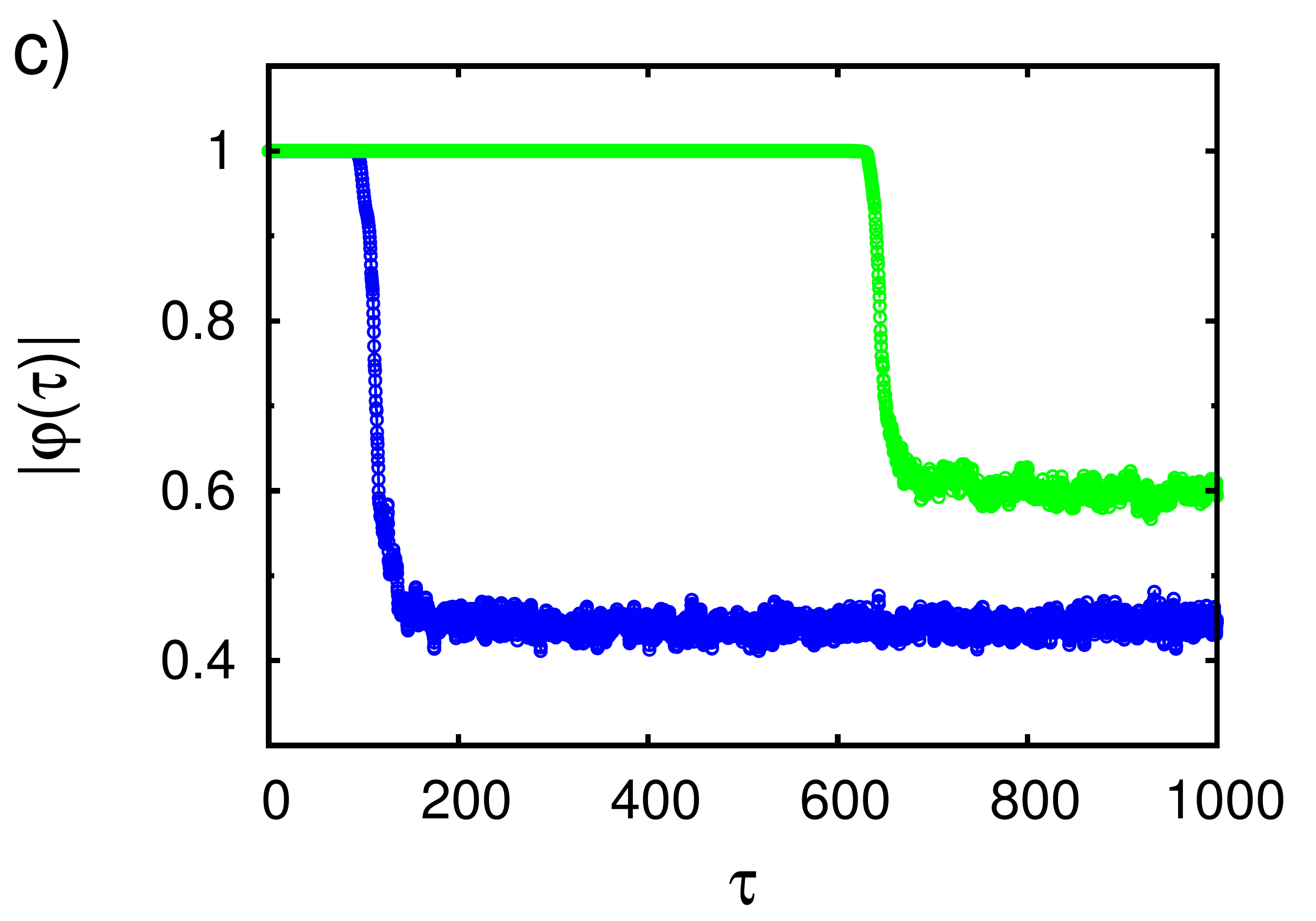}
\vspace{0.6cm}
\caption{(Color online) Panels (a) and (b) are contour plots of $\Gamma(k;q)$ (see caption of Fig.~\ref{fig1})
for $\bar{U}=\bar{V}=1$ and $\alpha=2.5$, $\beta=\infty$ (long-range
interaction) and $\alpha=\infty$, $\beta=2.5$ (long-range hopping).
In panel (c) we depict the time evolution of the order parameter 
defined in Eq. \eqref{order_parameter} for the initial conditions indicated in the
contour plots with points. As we can see the long-range hopping 
instability, upper (green) curve, takes a much longer time to set in 
than the long-range interaction instability, lower (blue) curve.
}
\label{fig8}
\end{figure}

To compare the time scales on which these two kind of 
instabilities act we have considered a case where 
we switch on and off alternatively the long-range interaction
and hopping (see Fig.~\ref{fig8}). We prepare the two systems with
a planewave $k$ slightly inside the instability region
perturbed by a $q$ equaling $\pi$ and $2 \pi / L$ (the widest available
perturbation with a nonvanishing imaginary Bogoliubov frequency).
By inspecting the contour plots [Figs.~\ref{fig8}(a) and (b)] we foresee a 
longer lifetime for the instability induced by the non-local hoppings. 
This result is confirmed by numerical simulations, shown 
in panel $c$ of Fig.~\ref{fig8}(c).
This is a general feature of the long-range hopping instability: we observe 
that at the the critical value $k_{\text{cr}}$, by definition, 
$\Gamma(k_{\text{cr}};q_{\text{max}})=0$ both for 
non-local interaction and hoppings. However, entering the unstable 
region gives a vanishing value of $\Gamma(k;q)$ for $q=0$ and $k>k_{\text{cr}}$  
and $\Gamma(k;q) \propto q$ for $q$ small for long-range hoppings. 
At variance for long-range interactions 
as $k$ is slightly larger than $k_{\text{cr}}$ then 
$\Gamma(k;q_{\text{inst}})$ generally acquires a finite value at the 
wavevector $q_{\text{inst}}$ at which the instability arises [with 
$q_{\text{inst}}=\pi$ for the non-competing case and $q_{\text{inst}}=q^\ast\in(0,\pi/2)$ 
for the competing one].

Let us examine the peculiarities of the instabilities
due to interaction and hopping and examine whether
these may be reproduced with finite-range couplings
or hoppings.
As far as the long-range interaction is concerned, in the non competing
case (examined in Sec.~\ref{MI_LR_noncompeting}),
we observe that the situation is not very different
from the finite-range one with nearest-neighbor interaction 
($\alpha \rightarrow \infty$). We can indeed find an 
effective nearest-neighbor $\tilde{\bar{V}}$ 
interaction giving rise to the same $k_{\text{cr}}$ [see Eq.~\eqref{k_cr_V}]:
\begin{equation}
 \tilde{\bar{V}}=\bar{V} \zeta(\alpha) (1-2^{1-\alpha}).
\end{equation}

\begin{figure}[t]
\includegraphics[width=.23\textwidth,angle=270]{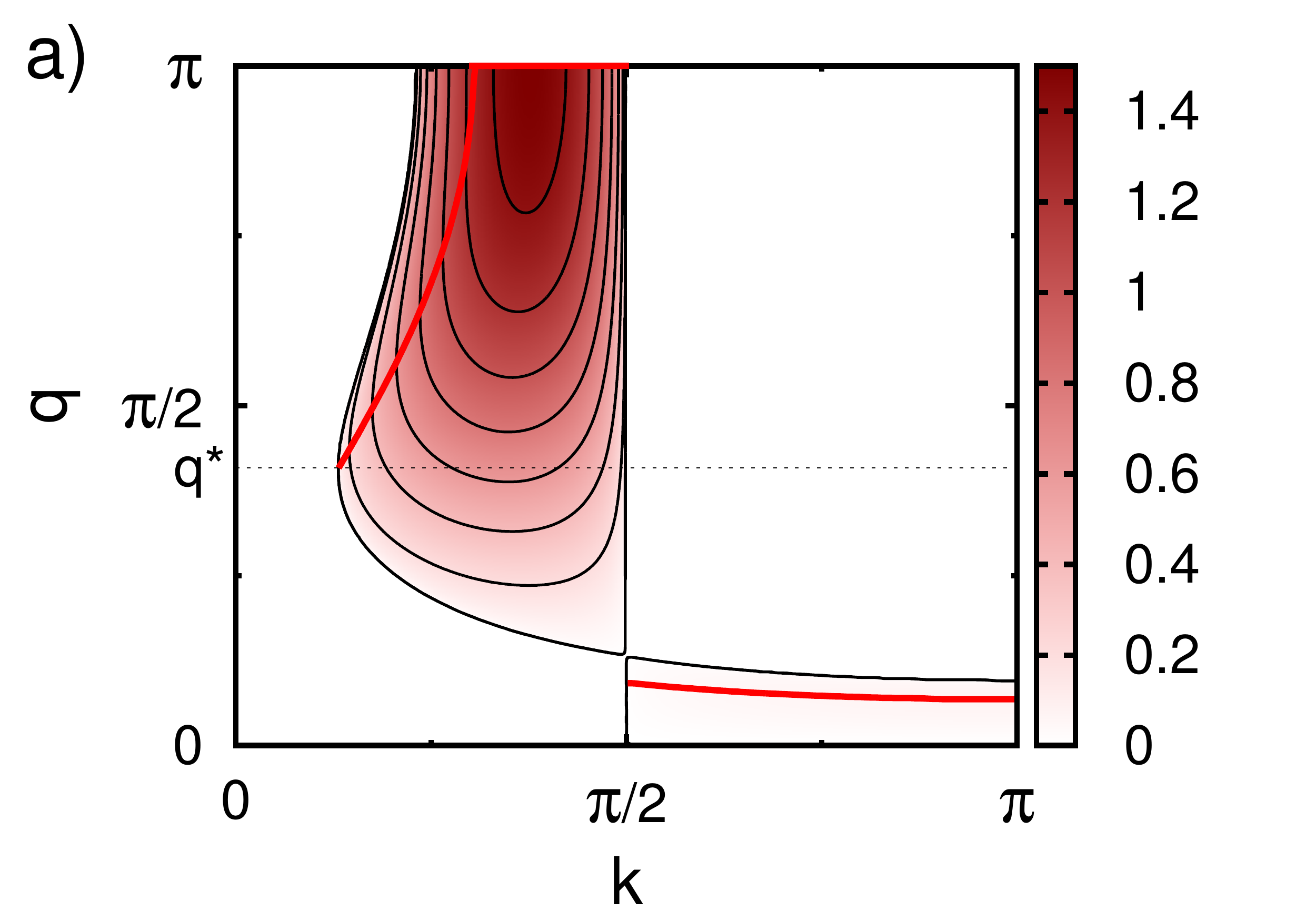}
\includegraphics[width=.23\textwidth,angle=270]{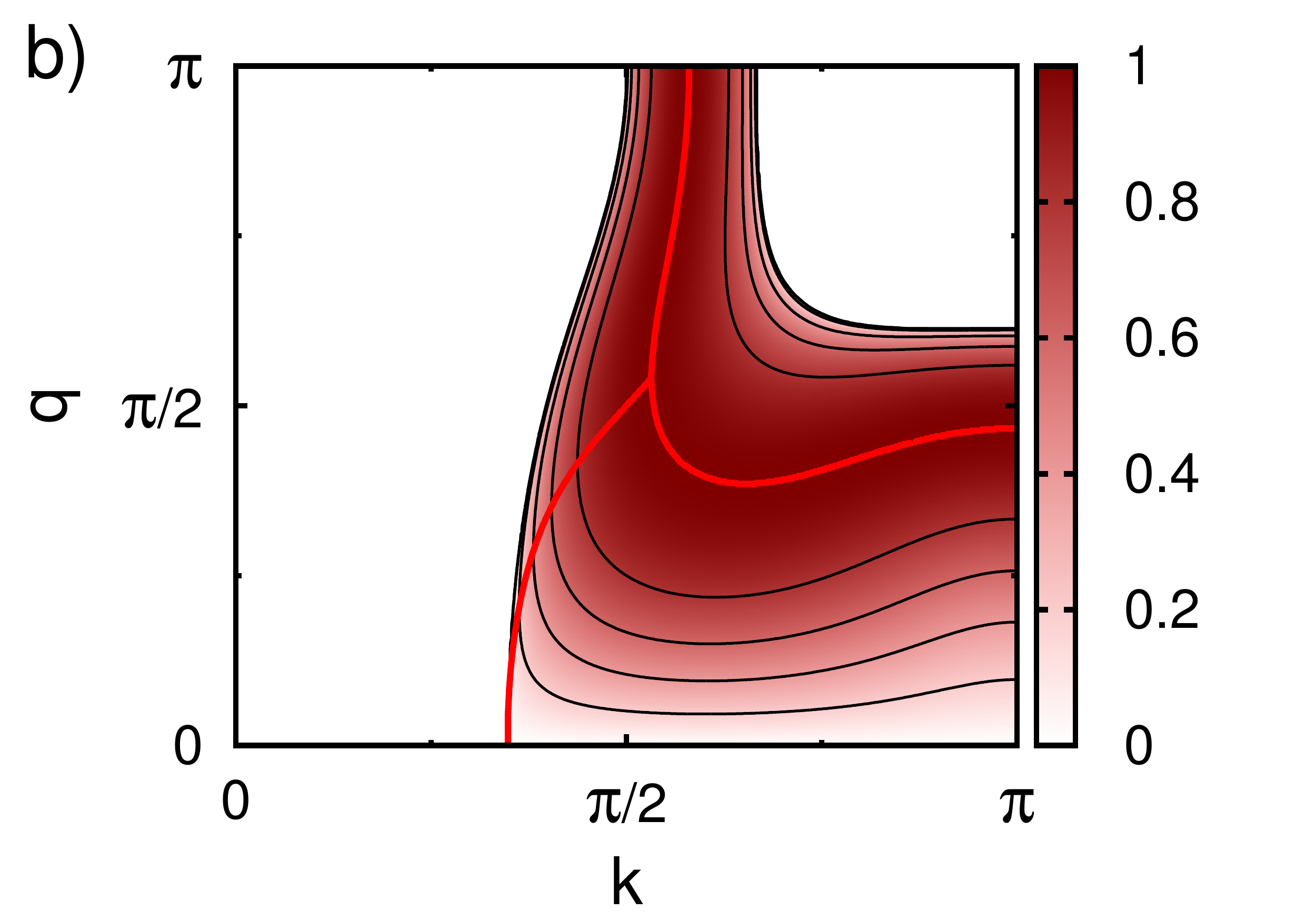}
\vspace{0.6cm}
\caption{(Color online) Contour plots of $\Gamma(k;q)$ (see caption of Fig.~\ref{fig1})
for models including only on-site interactions ($U$)
and next- and next-to-nearest-neighbor
interaction ($V$ and $V_{j,j+2}=V_2$) and hoppings 
($t=1$ and $t_{j,j+2}=t_2$), respectively. 
Panel (a) refers to values $U=-0.8$, $V=0.4$, 
$V_2=0.05$, and $t_2=0$ while
panel (b) refers to values $U=1$, $V=V_2=0$, and 
$t_2=0.2$.
}
\label{fig9}
\end{figure}

Similarly in the case where the competition in present (described in
Sec.~\ref{MI_LR_competing}) the case with $\alpha$ finite 
can be seen to be similar to the one obtained with $\alpha\rightarrow\infty$.
In fact one can generate a new wavevector $q^*$ even when
$\alpha\rightarrow\infty$, and thus the long-ranged-ness
of the interaction is not actually playing a major role.
As an example in Fig.~\ref{fig9}(a) we have 
considered a case where a nearest-neighbor and 
next-to-nearest-neighbor
interaction generates a stability diagram
closely resembling the one shown in Fig.~\ref{fig5}.

When we move to the hopping instabilities
(examined in Sec.~\ref{MI_LR_hopping})
we observe that while a finite-range hopping 
can generate $q=0$ instabilities
for $k<\pi$, the value of $k_{\text{cr}}$ is effectively
limited by the range of the hopping and it is not $0$.
This can be seen considering the hopping coefficients 
of range $R$, i.e., $t_{ij}=0$ for $\mid i-j \mid > R$. 
An example is given by $R=2$ where 
we set $t_{j,j \pm 1} \equiv t_1$, $t_{j,j \pm 2} \equiv t_2$. 
Explicit expressions (not reported here) for the stability 
regions can be derived repeating the analysis presented 
in Sec.~\ref{MI_analysis}. It is possible to show that for a hopping
of range $R$ then the critical momentum $k_{\text{cr}}$ can become as small as 
$\approx \pi/\left(2 R\right)$ due to the $q=0$ instability. 
Thus for $\beta>2$ one can find a finite-range hopping with range 
$R$ to reproduce the critical value $k_{\text{cr}}$ and the stability regions 
of non-local hoppings with exponent $\beta$, while this is not the case 
for $1<\beta<2$. We conclude that the case $1<\beta<2$ where $k_{\text{cr}}=0$ 
is singled out as a case where the instability is \emph{genuinely}
due to the long-range nature of the hoppings.
In Fig.~\ref{fig9}(b) we depict
the stability diagram of a case with the nearest-
and next-to-nearest-neighbor hoppings mimicking
the effect of long-range hopping shown in
Fig.~\ref{fig6}(a) and~\ref{fig6}(b), with a finite value 
of $k_{\text{cr}}$.

\section{Physical Applications}
\label{phys_app}

The cases we have analyzed can be applied to the 
study of modulational instability in a variety of physical 
systems characterized by non-local interaction and hopping coefficients. 
For example, the excitation transfer energy 
in molecular crystals and biopolymers 
exhibits a $1/r^3$ decay due to the dipole-dipole 
interaction \cite{davydov71,scott92}. The possibility of the inclusion 
of a non-local hopping is also present in  DNA modeling (e.g., in the 
the Peyrard-Bishop model \cite{peyrard89}), where the equation of motion
for the transverse stretching of the hydrogen bonds connecting the
bases facing each other contains a long-range hopping term
due to the dipole-dipole interaction among hydrogen bonds
resulting in a value $\beta=3$ \cite{cuevas02}. Another physical system 
characterized by non-local interactions 
is provided by ultracold atomic dipolar gases in optical 
lattices \cite{lahaye09,trefzger11}, for which experimental 
results recently appeared \cite{muller11,billy12,depaz12}.
For dipolar bosons as $^{52}\mathrm{Cr}$ in the 
Bose-Einstein condensate phase the low-energy effective Hamiltonian 
is expected to be the DNLSE in deep optical lattices 
with the power-law 
interaction coefficients having $\alpha=3$. 
We also mention that for ultracold bosons 
in suitably tailored optical lattices it is possible to 
have non-local hoppings $t_1$, $t_2$ \cite{greschner12}. 
To be specific, in the following we study the instability threshold for the
case of $^{52}\mathrm{Cr}$ trapped bosons in a quasi-one-dimensional geometry.
We provide estimates of the DNLSE parameters $t$, $U$, $V$ in terms of the 
physical parameters, showing 
that the critical value $k_{\text{cr}}$ does not crucially depend on the exponent $\alpha$. A similar 
computation is presented for a model having non-local hoppings $t_1$, $t_2$, 
as the one described in \cite{greschner12}. Also in this case
we find a relatively small quantitative effect on the value of the critical 
value $k_{\text{cr}}$. As discussed in Secs. \ref{MI_LR_interaction} to \ref{MI_interaction_hopping_comparison}, 
there is no qualitative difference 
between a finite value of $\alpha$ or $\beta$ and the corresponding results for $\alpha \to \infty$ or 
$\beta \to \infty$, as soon as that $\alpha>1$ or $\beta>2$. The results presented in this section 
further show that the quantitative difference 
(say, comparing $\alpha=3$ and $\alpha \to \infty$ results) is rather small.

The Gross-Pitaevskii Hamiltonian for a Bose-Einstein condensate 
of dipolar bosons in a quasi-one-dimensional trap in presence of 
an optical lattice is given by 
$$
H=\int dx \, \psi^*(x) \left( - \frac{\hbar^2}{2m} \frac{\partial^2}{\partial x^2} 
+ V_{lat}(x) + \frac{g_{1D}}{2} \mid \psi(x) \mid^2 \right) \psi(x) +
$$
\begin{equation}
+ \iint dx \, dy \, \psi^*(x) \psi^*(y) V_{1D}^{(dip)}\left(\mid x-y \mid \right) 
\psi(y) \psi(x).
\label{GPE_dip} 
\end{equation}
In Eq. \eqref{GPE_dip} $\psi(x)$ is the condensate wavefunction and 
$V_{\text{lat}}$ is the periodic potential due to the optical lattice along the $x$ 
direction, reading as $V_{\text{lat}}(x)=V_0 \cos^2{(kx)}$ where $k=2\pi/\lambda$ 
($\lambda/2$ is the lattice spacing). The strength $V_0$ is usually measured 
in units of the recoil energy $E_R=\hbar^2 k^2/2m$: We set $V_0\equiv s E_R$. 
The condensate wavefunction is normalized to the total number of particles 
$N_T$ and the filling $\rho$ is defined by $\rho=N_T/N_W$, where $N_W$ 
is the number of wells. Moreover, in Eq. \eqref{GPE_dip} $g_{\text{1D}}$ is the 
effective one-dimensional coupling constant, which is determined in terms of the 
3D $s$-wave scattering length $a$ and the size of the transverse 
confinement $\ell$ \cite{olshanii98}, which in turn depends on the radial 
confinement frequency $\omega_\perp$ (we will consider values of $\ell/a$ 
much smaller than one, far from the confinement-induced resonance). 
The dipole-dipole interaction induces a non-local two-body potential 
decaying as $1/r^3$. The dipole-dipole scattering length is defined as 
$a_{dd}=\mu_0 d^2 m/12\pi\hbar^2$, where $d$ is the dipole moment and 
$\mu_0$ is the vacuum permeability \cite{lahaye09}.
For dipolar gases in quasi-one-dimensional 
geometries it is possible to obtain an expression 
in the single-mode approximation for the effective non-local 
interaction $V_{\text{1D}}^{\text{(dip)}}$ by integrating the transverse 
ground-state \cite{sinha07}: one obtains 
$V_{\text{1D}}^{\text{(dip)}}(x)=(2 \alpha_{\text{orient}} d^2/\ell^3) [2 \sqrt{t}-\pi (1+2 t) e^t 
\mathrm{erfc}(\sqrt{t})]$, where $t=x/\ell$ and 
$\mathrm{erfc}$ is the complementary error function; the dimensionless constant 
$\alpha_{\text{orient}}$ depends on the angle $\varphi_{\text{orient}}$ the dipoles form with 
the $x$ axis and it may vary between $1$ (corresponding to $\varphi_{\text{orient}}=0$) and 
$-1/2$ ($\varphi_{\text{orient}}=\pi/2$). 

The DNLSE is found for deep lattices by using a tight-binding ansatz 
\cite{trombettoni01,morsch06} of the form 
$\psi(x,t)=\sum_j \Phi_j (x) \psi_j(t)$ where $\Phi_j$ is the Wannier 
function centered in the $j$-th well (and assumed in the following 
estimates to have a shape independent of $j$). The coefficients $t$, $U$ and 
$V$ are expressed as suitable overlap integrals of Wannier functions and 
can be estimated by using a Gaussian form for the $\Phi$'s, with the 
width being a parameter to be variationally determined (see the discussions 
in \cite{vanoosten01,trombettoni05}). We mention that the effect of 
the inter-site interaction term $V$ in the Bose-Hubbard phase 
diagram was recently investigated 
in \cite{dallatorre,berg08,amico10,dalmonte11,rossini12,giuliano13}.

The critical momentum $k_\text{cr}$ is an important quantity which has
been theoretically investigated
in \cite{smerzi02} and experimentally detected in \cite{cataliotti03}
for a system with short-range interactions.
We recall that experimentally the modulational instability may be
triggered by subjecting the system to
a sudden shift of the optical lattice or of confining harmonic trap
(as done in \cite{cataliotti03}) .\\

\begin{figure}[t]
\includegraphics[width=.5\textwidth,angle=270]{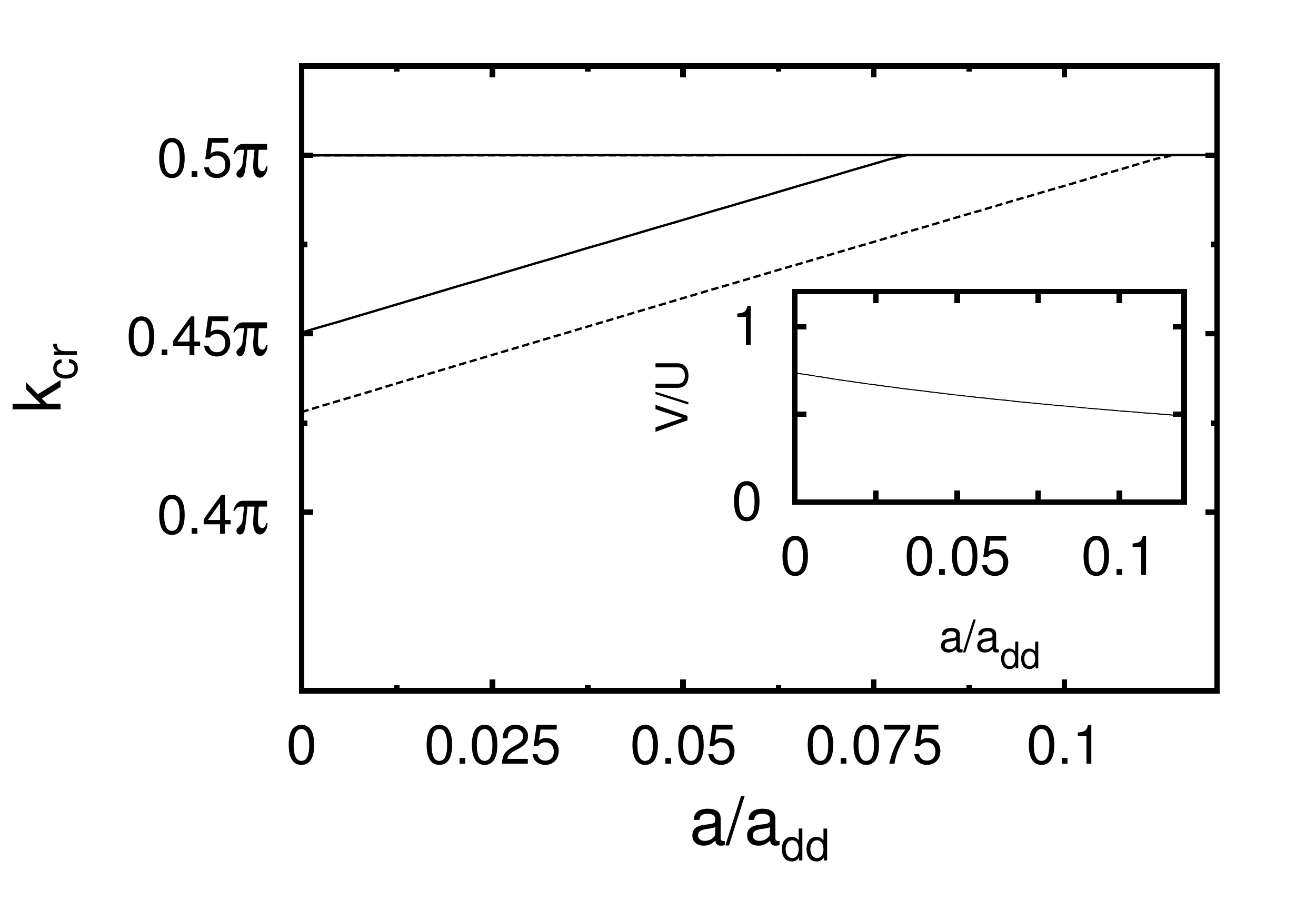}
\vspace{0.6cm}
\caption{$k_{\text{cr}}$ vs $a/a_{dd}$ for a dipolar gas of $^{52}\mathrm{Cr}$ atoms. 
Solid lines refer to $\alpha=3$ (top solid line: $\rho=1$, bottom solid line: $\rho=100$), 
while dotted lines refer to $\alpha \to \infty$ 
(top dashed line: $\rho=1$, bottom dashed line: $\rho=100$). 
Parameters are $\omega_\perp=2 \pi \, 100\text{Hz}$, $s=5$, 
$\alpha_{\text{orient}}=-1/2$ and $\lambda=0.7 \mu\text{m}$. Inset: Corresponding value of 
$V/U$ vs $a/a_{dd}$ for $\rho=1$, solid line, and $\rho=100$, dotted line 
(the two lines practically coinciding with one another).}
\label{fig10}
\end{figure}

\begin{figure}[t]
\includegraphics[width=.5\textwidth,angle=270]{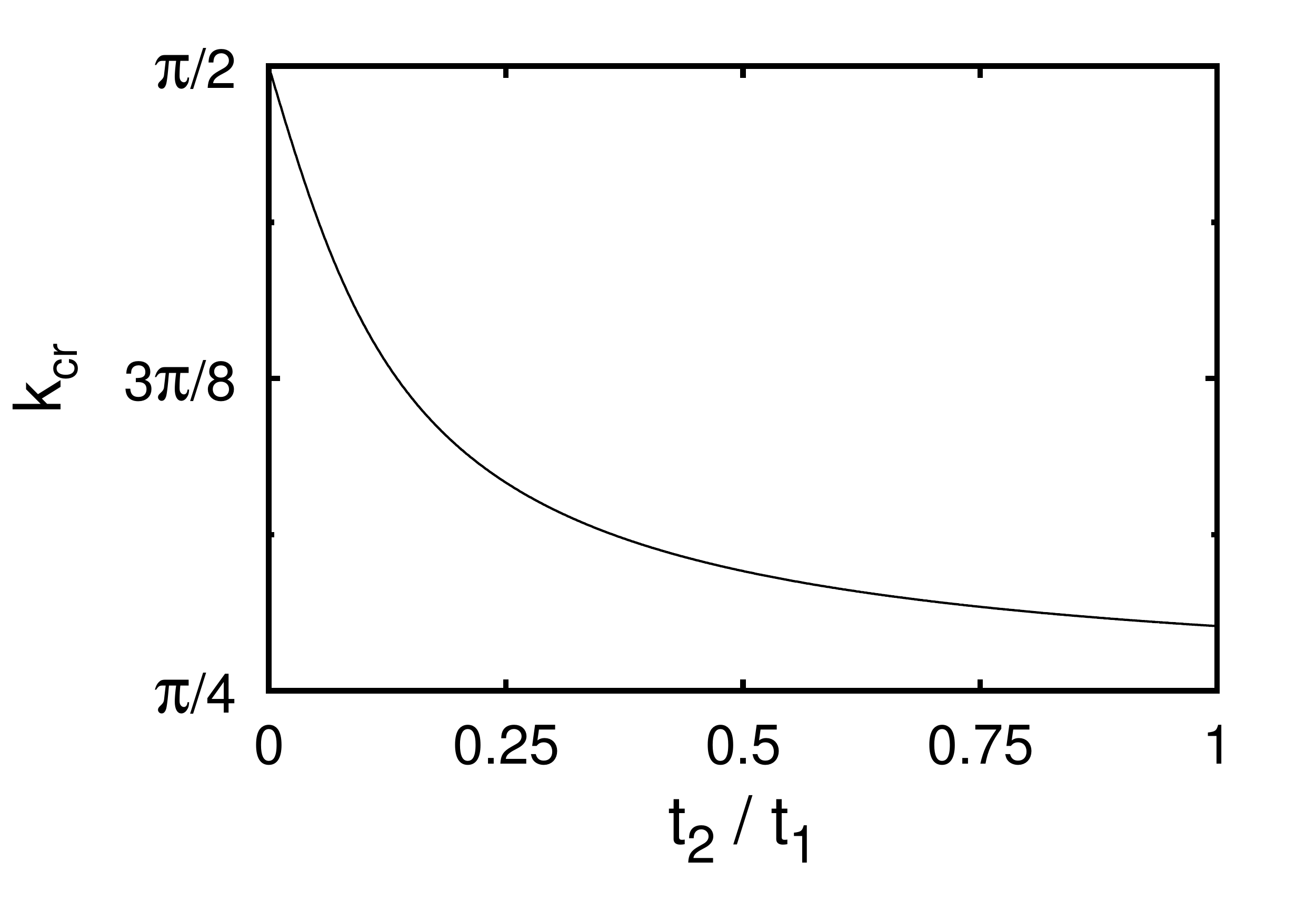}
\vspace{0.6cm}
\caption{$k_{\text{cr}}$ vs $t_2/t_1$ for a model 
having nearest-neighbor and next-nearest-neighbor hoppings 
$t_1$ and $t_2$ ($V=0$).}
\label{fig11}
\end{figure}

Our results for the critical value 
$k_{\text{cr}}$, after computing the parameters 
$t$, $U$, and $V$, are drawn in Fig.~\ref{fig10} 
for a set of typical experimental parameters: we fixed 
$\omega_\perp=2 \pi \, 100\text{Hz}$, $s=5$, 
$\alpha_{\text{orient}}=-1/2$ (corresponding to repulsion) and 
$\lambda=0.7 \mu \text{m}$ and we varied 
$a/a_{dd}$ for $^{52}\mathrm{Cr}$ atoms for two different values of the filling 
$\rho$ (we choose $\rho=1$ and $\rho=100$). In the inset of Fig.~\ref{fig10} 
we plot the ratio $U/V$ versus $a/a_{dd}$ to quantify how much the interaction 
is non-local for a typical value of the parameters. One sees that deviations 
are observed from the critical value $k_{\text{cr}}=\pi/2$ obtained without non-local 
interactions and these deviations are not crucially dependent on $\alpha$. 
In Fig.~\ref{fig11} we finally plot the critical value $k_{\text{cr}}$ for a model 
having $t_1$ and $t_2$ (i.e., $R=2$) with $V=0$: The critical value $k_{\text{cr}}$ 
does not depend on $U$ and smoothly passes from $\pi/2$ (for $t_2=0$) 
to $\pi/4$ (for $t_1=0$).

\section{Conclusions}
\label{conclusions}

In this paper we studied 
the occurrence of modulational instabilities in nonlinear lattices 
with long-range hoppings and interactions. We were motivated 
by experiments of (di)polar gases in optical lattices and by the 
interest in the study of dynamical regimes in systems with long-range 
interactions. Using the discrete nonlinear Schr\"odinger equation in 
one dimension, we considered 
power-law decaying interactions (with exponent $\alpha$) 
and hoppings (with exponent $\beta$). 
We showed that the effect of long-range interactions 
is that of shifting the onset of the modulational instability 
region for $\alpha>1$ (corresponding to an extensive 
energy). At a critical value of the interaction strength, 
the modulational stable region shrinks to zero. Similar results are found 
for short-range non-local hoppings ($\beta>2$). At variance, 
for longer-ranged hoppings ($1<\beta<2$) 
there is no longer any modulational stability. Explicit estimates 
for the critical 
values of the momentum at which the system becomes unstable are presented 
for a quasi-one-dimensional ultracold dipolar gas in a deep optical lattice.

Instabilities due to the interaction generally differ
from the ones due to hopping since the 
first ones are sensitive to finite wavelength perturbations
while the second ones are most sensitive to perturbation
of the order of the system size. 
Such hopping generated instability turns out to have 
generally longer timescales than the interaction
generated ones.
If we allow interactions acting on different scales to compete
we may generate instabilities with longer, but finite, wavelengths,
in analogy with what is met in other systems with competing 
interactions \cite{leshouches10}.

The instabilities met in the long-range interacting and 
long-range hopping for $\beta>2$ are not specific to the 
long-ranged nature of the interaction or hopping. 
In fact, their effects are, in principle, not different from
finite range cases.
As far as very long-ranged ($1<\beta<2$) hopping is concerned,
we found that it gives rise to \emph{genuinely} long-range 
instabilities, since they cannot be reproduced with suitably chosen 
finite-range hoppings.

\textit{Acknowledgements} 
We thank M. Iazzi, A. Smerzi, G. De Ninno 
and F. Staniscia for very useful discussions. 

\appendix

\section{Useful properties of $\ell_\alpha(k)$}
\label{appendix_A}

\begin{figure}[t]
\centerline{\includegraphics[width=.5\textwidth,angle=270]{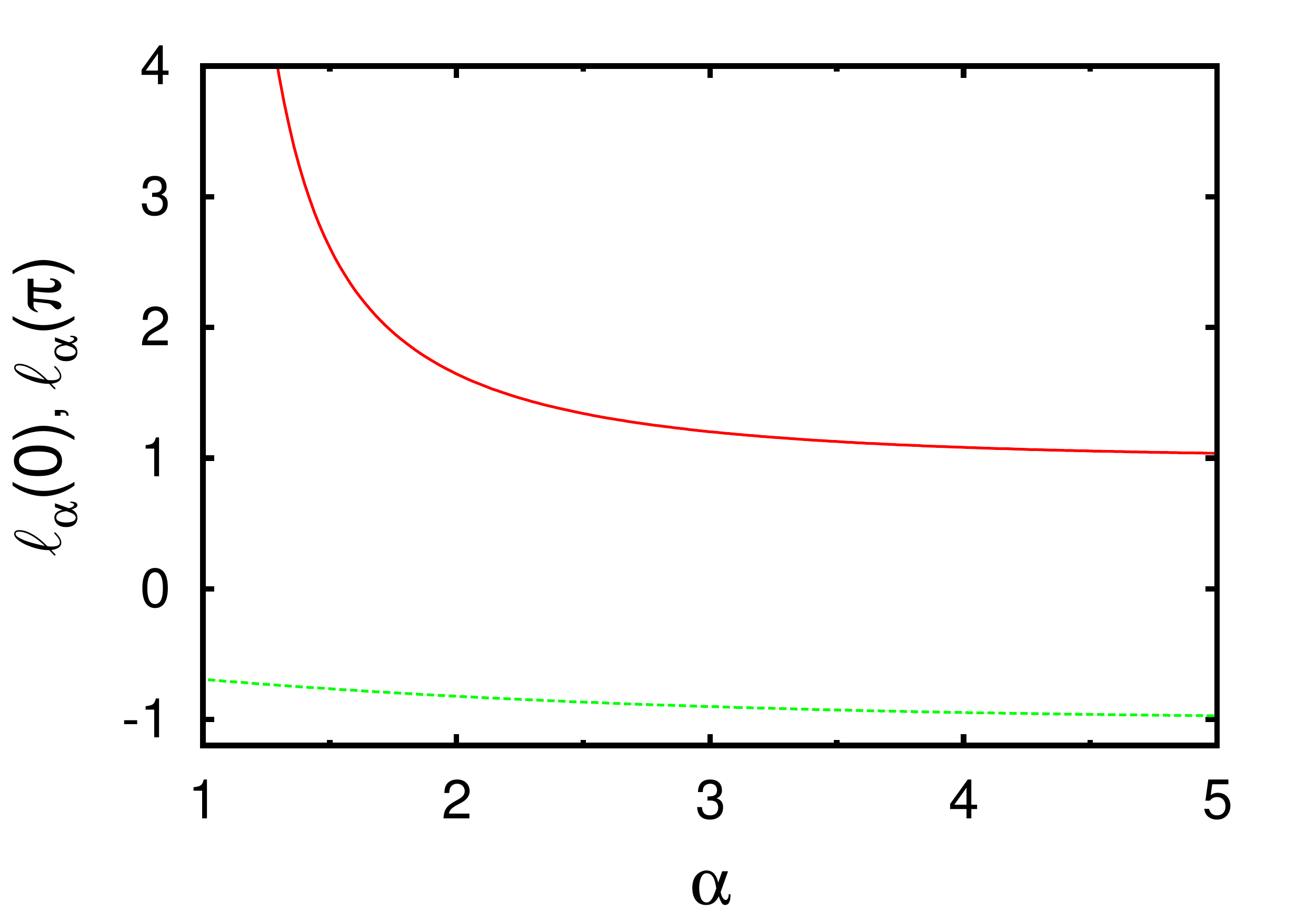}}
\vspace{0.6cm}
\caption{(Color online) Plot of $\ell_\alpha(0)$ [solid (red) line] and 
$\ell_\alpha(\pi)$ [dashed (green) line]
as a function of $\alpha$.}
\label{app_1}
\end{figure}

The analysis of the stability regions presented in the main text is based on the 
study of the quantity ${\cal I}$ defined in \eqref{II}, 
which in turn contains the function $\ell_\alpha(q)$ defined in Eq.\eqref{elle_funct}:
\begin{equation}
\ell_\alpha(q)= \sum_{m=1}^{\infty} \frac{\cos{(mq)}}{m^\alpha}
\label{elle_funct_app}
\end{equation}
(with $\alpha>1$). We are interested in the domain $q \in [0,\pi]$. 
From the definition it follows that $\ell_\alpha(0)=\zeta(\alpha)$; it is also 
\begin{equation}
\ell_\alpha(\pi)=- \left( 1-2^{1-\alpha}\right) \, \zeta(\alpha).
\end{equation}
The plot of $\ell_\alpha(0)$ and $\ell_\alpha(\pi)$ as a function of $\alpha$ 
is drawn in Fig.~\ref{app_1}. Notice that the behavior of $\ell_\alpha(0)$ 
for $\alpha \to 1$ and $\alpha \to \infty$ is given respectively 
by $\lim_{\alpha \to 1} \ell_{\alpha}(0)=\infty$ and 
$\lim_{\alpha \to \infty} \ell_{\alpha}(0)=1$. For $\ell_\alpha(\pi)$ one has 
$\lim_{\alpha \to 1} \ell_{\alpha}(\pi)=\ln{2}<0$ and 
$\lim_{\alpha \to \infty} \ell_{\alpha}(\pi)=-1$.

The derivative of $\ell_{\alpha}$ has a different behavior for $1<\alpha<2$, 
$\alpha=2$ and $\alpha>2$. It is
$$
\frac{\partial \ell_\alpha}{\partial q}\left( \pi \right)=0
$$
for $\alpha>1$ and 
\begin{equation}
\frac{\partial \ell_\alpha}{\partial q}\left( 0 \right)=
\Bigg\{ \begin{array}{cc}
0 & \alpha>2\\
-\frac{\pi}{2} & \alpha=2\\
-\infty & 1<\alpha<2 \end{array} 
\end{equation}

The second derivative of $\ell(\alpha)$ can be computed explicitly and it gives:
\begin{equation}
\frac{\partial^2 \ell_\alpha}{\partial q^2}\left( q \right) = - \ell_{\alpha-2} (q).
\end{equation}
For $1<\alpha \le2$ then $\frac{\partial^2 \ell_\alpha}{\partial q^2}\left( q \right)$
is a positive function and we have 
$\frac{\partial^2 \ell_2}{\partial q^2}\left( q \right) = \frac{1}{2}$,
constant over $q\in [0,\pi]$.
The second derivative takes the following values at the extrema of the 
interval $\left[0,\pi \right]$ for $\alpha>1$:
\begin{equation}
\frac{\partial^2 \ell_\alpha}{\partial q^2}\left( 0 \right)=
\left\{ \begin{array}{cc}
\infty & \alpha<2\\
1/2 & \alpha=2\\
-\infty & 2<\alpha\le3\\
-\zeta(\alpha-2) & \alpha>3. \end{array}
\right. 
\end{equation}
Finally at $q=\pi$ we have $\frac{\partial^2 \ell_\alpha}{\partial q^2}\left( \pi \right) 
= (1-2^{3-\alpha})\zeta(\alpha)$ for every $\alpha>1$.

\end{document}